\documentclass[aps,prx,reprint,nofootinbib,preprintnumbers,floatfix,unsortedaddress,
superscriptaddress,longbibliography]{revtex4-2}
\expandafter\let\csname ver@array.sty\endcsname\relax
\usepackage{array}

\input{header-1}

\newcommand{\bs}{{\mathbf{s}}}

\newcommand{\ua}{{\uparrow}}
\newcommand{\da}{{\downarrow}}

\newcommand{\dproj}{{d_{\mathrm{proj}}}}
\newcommand{\projS}{\mathbb{P}}
\newcommand{\dprojres}{{d_{\mathrm{proj}}^{\mathrm{res}}}}

\newcommand{\nn}{{\langle ij\rangle}}
\newcommand{\nnn}{{\langle\!\langle ij\rangle \!\rangle}}

\renewcommand\vec\mathbf

\newcommand\redsout{\bgroup\markoverwith{\textcolor{red}{\rule[0.5ex]{2pt}{1.4pt}}}\ULon}

\usepackage{relsize}

\usepackage[acronyms, nohypertypes={acronym}, nopostdot, style=super, nonumberlist, toc]{glossaries}
\setacronymstyle{long-sc-short}
\glsenableentrycount
\makeglossaries

\newcommand\ac[1]{\gls{#1}}

\newacronym{PNA}{pna}{particle-number-algorithm}
\newacronym{SFA}{sfa}{spin-flip-algorithm}

\newacronym{WF}{wf}{Wilson-Fisher}
\newacronym{AF}{af}{asymptotically free}

\newacronym{RG}{rg}{renormalization group}

\newacronym{SSM}{ssm}{Shastry-Sutherland Model}

\newacronym{TFIM}{tfim}{Transverse Field Ising Model}

\newacronym{QIS}{qis}{Quantum Information Science}
\newacronym{PPT}{ppt}{positive-semidefinite partial transpose}
\newacronym{NPT}{npt}{negative partial transpose}

\newacronym[longplural={conformal field theories}]{CFT}{cft}{conformal field theory}
\newacronym[longplural={lattice field theories}]{LFT}{lft}{lattice field theory}
\newacronym[longplural={effective field theories}]{EFT}{eft}{effective field theory}
\newacronym[longplural={quantum field theories}]{QFT}{qft}{quantum field theory}

\newacronym[]{DMRG}{dmrg}{Density Matrix Renormalization Group}
\newacronym[]{PDMS}{pdms}{Projected Density Matrix Sampling}

\newacronym[]{LOCC}{locc}{Local Operations and Classical Communicaton}
\newacronym[]{OBC}{obc}{open boundary conditions}

\newacronym{MPS}{mps}{matrix product states}

\newacronym{JLP}{jlp}{Jordan-Lee-Preskill}

\newacronym{BBN}{bbn}{big bang nucleosynthesis}

\newacronym{LEC}{lec}{low-energy constant}

\newacronym{QCD}{qcd}{quantum chromodynamics}
\newacronym{MC}{mc}{Monte Carlo}

\newacronym{IR}{ir}{infrared}
\newacronym{UV}{uv}{ultraviolet}

\newacronym{QED}{qed}{quantum electrodynamics}
\newacronym{SNR}{snr}{signal-to-noise ratio}

\newacronym{NLSM}{nlsm}{nonlinear sigma model}

\newacronym{CL}{cl}{Complex Langevin}

\newacronym{CSA}{csa}{Cartan subalgebra}

\newacronym{SSB}{ssb}{spontaneous symmetry breaking}

\newacronym{AFQMC}{afqmc}{auxiliary field quantum Monte Carlo}
\newacronym{iHMC}{ihmc}{imaginary-mass Hybrid Monte Carlo}

\newacronym{MCMC}{mcmc}{Markov Chain Monte Carlo}

\newacronym{QI}{qi}{quantum information}

\begin{document}

\title{Projected Density Matrix Sampling for Lattice Hamiltonians}
\author{Abhishek Karna \orcidlink{0009-0007-4005-5179}}
\email{abhishek.karna@duke.edu}
\affiliation{ Department of Physics, Duke University, Box 90305, Durham, North Carolina 27708, USA}
\author{Hansen S. Wu \orcidlink{0009-0003-0012-9082}}
\email{hsw5129@psu.edu}
\affiliation{Department of Physics, Pennsylvania State University, State College, Pennsylvania 16801, USA}
\author{Shailesh Chandrasekharan\orcidlink{0000-0002-3711-4998}}
\email{sch27@duke.edu}
\affiliation{ Department of Physics, Duke University, Box 90305, Durham, North Carolina 27708, USA}
\author{Ribhu K. Kaul\orcidlink{0000-0003-1301-7744}}
\email{ribhu.kaul@psu.edu}
\affiliation{Department of Physics, Pennsylvania State University, State College, Pennsylvania 16801, USA}

\date{\today}

\begin{abstract}
Quantum Monte Carlo methods are powerful tools for studying quantum many-body systems but face difficulties in accessing excited states and in treating sign problems. We present a continuous-time path-integral Monte Carlo method for computing the low-lying spectrum of generic quantum Hamiltonians within a projection subspace. The method projects the thermal density matrix onto a subspace spanned by a chosen set of linearly independent states. It is free of Trotter discretization errors and systematically converges to the low-energy states which have finite overlap with the projection subspace as the $\beta$ parameter increases. While most effective for systems without a sign problem, the method also yields information about low-energy spectra when sign problems are present. We illustrate the approach on two problems. For the sign-free case, we compute the first four low-energy levels in the scaling limit of the one-dimensional Ising model with both transverse and longitudinal fields, demonstrating the flow from the conformal limit to the massive $E_8$ quantum field theory. For the sign-problem case, we apply the method to the frustrated Shastry--Sutherland model and benchmark it against exact diagonalization on small lattices. We also present results for larger systems beyond the lattice sizes accessible to exact diagonalization, while limited to small $\beta$ where sign problems occur. Our method provides a general route toward quantum Monte Carlo spectroscopy for lattice Hamiltonians.
\end{abstract}

\maketitle

\section{Introduction}
\label{sec1}

Computing the low-lying spectrum of quantum Hamiltonians is a central challenge in many-body physics and remains an active area of research across fields ranging from condensed matter to nuclear and particle physics. With the recent surge of interest in using quantum computers to solve generic quantum problems, alternative classical methods that can benchmark and cross-check quantum results are particularly valuable. While tensor networks are the main tools used today for such cross-verification \cite{BanulsReview23}, Monte Carlo sampling also provides a powerful framework for addressing these problems. However, its applicability is often limited by the notorious sign problem, which, in its most general form, is computationally hard \cite{TroyerWiese2005}. The severity of the sign problem depends on several factors, including the details of the interactions, the choice of Hilbert-space basis used for Monte Carlo sampling, and the energy subspace being accessed. A variety of strategies have been explored to mitigate it in specific cases \cite{Iglovikov2015,Hangleiter2019,Pan:2022fgf}.

The fundamental difficulty arises because the full Hilbert space of a quantum system typically grows exponentially with system size which makes it hard to efficiently isolate the relevant low-energy subspace. A common approach begins with the thermal density matrix, $\rho = e^{-\beta H}$, which naturally suppresses high-energy contributions and projects onto the low-energy sector as $\beta$ becomes large. In general, there is no simple basis $[s]$ in which all matrix elements $\rho_{s_f,s_i} = \langle s_f| e^{-\beta H} |s_i\rangle$ can be expressed as a sum over positive weights, so that Monte Carlo sampling becomes possible. This is the origin of the sign problem.

A related idea that has not been explored as extensively, and which serves as the main motivation for this work, is to project the thermal density matrix onto a carefully chosen $\dproj$-dimensional {\it projection subspace} $\projS$, spanned by a set of linearly independent {\it projection states} $|\psi_a\rangle$ ($a = 1, 2, \ldots, \dproj$) within the Hilbert space. We then define the $\dproj \times \dproj$ matrix-valued partition function $Z$ as the {\it projected thermal density matrix}, whose elements are given by
\begin{align}
Z_{ac} &= \langle \psi_a | e^{-\beta H} | \psi_c \rangle.
\label{eq:Zmat}
\end{align}
Similarly, we define the corresponding {\it projected thermal energy matrix} $E$ in the same subspace, with matrix elements
\begin{align}
E_{ac} &= \langle \psi_a | H e^{-\beta H} | \psi_c \rangle.
\label{eq:Emat}
\end{align}
The $\dproj$ eigenvalues $e_a(\beta)$ of the matrix product $E Z^{-1}$ are referred to as the {\it projected thermal energies} of $H$, which depend on the chosen projection subspace.

The thermal energies $e_a(\beta)$ depend on the inverse temperature $\beta$ and converge, in the zero temperature limit, to what we refer to as the {\it projected low-energy spectrum} of $H$. It can be shown, using standard linear-algebraic arguments, that this spectrum corresponds to the $\dproj$ lowest eigenvalues of $H$ whose eigenstates not only have nonzero overlap with the projection subspace but also remain linearly independent after being projected onto it. 

This broad idea of extracting the low-energy spectrum using Monte Carlo techniques is well known in the lattice field theory community and is often discussed within the frame work of the generalized eigenvalue problem~\cite{Blossier:2009kd}. It is extensively employed to determine low-energy particle spectra \cite{Luscher:1990ck,DeGrand:2006zz,Esposito:2016noz,Liu:2019zoy}. Similar approaches have been proposed for constructing the reduced density matrix through a spatial bipartition for use in Monte Carlo simulations, and applied to compute entanglement entropy~\cite{PhysRevB.82.100409} and to extract entanglement level spectra in model quantum systems~\cite{PhysRevB.89.195147,mao2025sampling}. However, its applicability to a broad range of problems—particularly in condensed matter physics and in the Hamiltonian framework in the above form—has not yet been systematically explored. For instance, if the projection subspace coincides exactly with an energy subspace of the Hamiltonian, accurate results for the low-energy spectrum can be obtained using efficient algorithms even for arbitrarily small values of $\beta$, irrespective of whether the problem suffers from a sign problem \cite{Chandrasekharan:2024iao}. Furthermore, by judiciously enlarging the projection subspace, the $\beta$-dependence of the extracted energies can be exponentially suppressed when accurate Monte Carlo sampling is feasible \cite{Blossier:2009kd}. These insights, however, have not yet been effectively applied widely to physically interesting systems, like quantum spin models and qubit regularized Hamiltonians~\cite{Chandrasekharan:2025Cb}.

The goal of this work is to systematically investigate and develop the idea of sampling the thermal density matrix in a projection subspace in substantial detail. To that end, we introduce the \ac{PDMS} method with the aim of computing the projected thermal energies $e_a(\beta)$ for generic quantum Hamiltonians. Our algorithm samples configurations that contribute to the density matrix, expressed in the path-integral representation, directly in continuous time, thereby eliminating errors typically introduced by Trotter decompositions. All matrix elements of $Z$ and $E$ are computed as observables in the Monte Carlo procedure, up to an overall normalization factor that cancels when evaluating $E Z^{-1}$ to extract $e_a(\beta)$. Sign problems, if present, are incorporated into the observables in the usual manner.

Our work synthesizes several established ideas, including continuous-time path-integral formulations \cite{Beard:1996wj,RevModPhys.83.349,PhysRevB.102.094101}, the stochastic series expansion \cite{Sandvik1999_SSE,Sandvik2010_QMCReview,Sandvik2019_SSEReview}, worm and loop algorithms \cite{PhysRevLett.87.160601,Evertz:2000rk,Chandrasekharan:2024iao}, and meron-cluster techniques \cite{Chandrasekharan:1999cm}. To illustrate the method, we apply it to two representative problems: (1) the one-dimensional \ac{TFIM} in the presence of a small longitudinal magnetic field. Here, we demonstrate how our method can compute the low-lying mass spectrum of the $E_8$ quantum field theory that is known to emerge in this regime \cite{Zamolodchikov:1989fp}.  And, (2) the Shastry--Sutherland model (SSM)~\cite{ShastrySutherland1981}, a paradigmatic frustrated antiferromagnetic system with a severe sign problem but of great interest in condensed matter physics, owing to its exotic phase diagram and experimental relevance (see Ref.~\cite{Corboz2025} and references therein). In this case, our goal is to investigate how well our algorithm computes the values of $e_a(\beta)$ on small lattices for selected projection subspaces. These thermal energies can be obtained independently using exact diagonalization for comparison. We hope that these two distinct applications of our method, along with previous developments in similar directions, help convey its broad applicability across nuclear, particle, and condensed matter physics.

The rest of the paper is organized as follows. In \cref{sec2}, we describe our general formalism of representing the thermal density matrix in a continuous-time path-integral form and discuss the bookkeeping of the signs. In \cref{sec3}, we detail the two update algorithms for our Monte Carlo: (i) the link update and (ii) the spin update. Various details about leveraging the properties of different configuration representations and computation of the $\dproj \times \dproj$ projected thermal matrices $Z$ and $E$ are also discussed therein. As can be anticipated, the role of the choice of the projection subspace is crucial in our approach. We discuss this in \cref{sec4} by providing a linear algebraic description of the algorithmic process towards convergence to the low-energy spectrum for a generic quantum Hamiltonian $H$. We also comment on rates of convergence and potential solutions to the complications associated in models with sign problems. In \cref{sec5}, we apply our algorithm to the 1D-TFIM at criticality in the presence of magnetic field and obtain the first two predicted mass ratios of the associated $E_8$ massive quantum field theory. Similarly, in \cref{sec6}, we explore the 2D-frustrated SSM Hamiltonian and demonstrate the effectiveness of different projection subspaces at different coupling ratios. We also discuss strategies for improving the convergence by either enlarging the dimension of the projection subspace $\dproj$ or choosing an optimal projection basis, thereby partially mitigating the sign problem. We summarize our discussions in \cref{sec7}.

\section{Thermal Density Matrix in Continuous Time}
\label{sec2}

To construct a Monte Carlo sampling procedure for the thermal density matrix, we formulate it in a continuous-time path-integral representation.\footnote{While we formulate the algorithm in the path integral representation, our ideas for computing the $Z$ and $H$ matrices and using them to extract the low lying states can clearly equally well be formulated in the popular stochastic series expansion representation~\cite{Sandvik1999_SSE}. This is outlined in App.~\ref{app:sse}.} We focus on Hamiltonians that can be written in the generic form
\begin{align}
H = \sum_{\ell} H_\ell + E_0,
\label{eq:mcH}
\end{align}
where $\ell$ labels a bond associated with local interactions, and $H_\ell$ denotes the Hamiltonian corresponding to the interaction on that bond. In the two applications considered in this work, $\ell$ corresponds to interactions between two lattice sites; however, in general, $\ell$ can involve interactions among multiple sites, such as those encountered in lattice gauge theories. The constant $E_0$ is chosen so that the diagonal matrix elements of $H_\ell$ are zero or negative in the chosen basis.

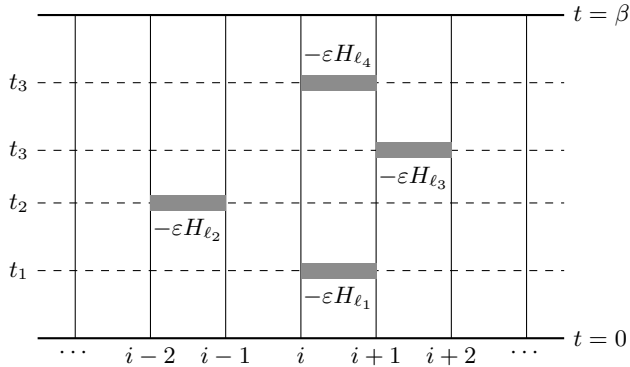
\begin{figure}[ht]
\centering
\begin{tikzpicture}

\draw[thick] (-0.5,0) -- (6.5,0) node[right] {$t=0$};
\draw[thick] (-0.5,4.3) -- (6.5,4.3) node[right] {$t=\beta$};

\foreach \x in {0,1,2,3,4,5,6}
    \draw[thin] (\x,0) -- (\x,4.3);

\node[below] at (0,0) {$\cdots$};
\node[below] at (1,0) {$i-2$};
\node[below] at (2,0) {$i-1$};
\node[below] at (3,0) {$i$};
\node[below] at (4,0) {$i+1$};
\node[below] at (5,0) {$i+2$};
\node[below] at (6,0) {$\cdots$};

\draw[dashed] (-0.5,0.9) -- (6.5,0.9);
\node[left] at (-0.5,0.9) {$t_1$};

\fill[gray!90] (3,0.8) rectangle (4,1.0);
\node[below] at (3.5,0.8) {$-\varepsilon H_{\ell_1}$};

\draw[dashed] (-0.5,1.8) -- (6.5,1.8);
\node[left] at (-0.5,1.8) {$t_2$};

\fill[gray!90] (1,1.7) rectangle (2,1.9); 
\node[below] at (1.5,1.7) {$-\varepsilon H_{\ell_2}$};

\draw[dashed] (-0.5,2.5) -- (6.5,2.5);
\node[left] at (-0.5,2.5) {$t_3$};

\fill[gray!90] (4,2.4) rectangle (5,2.6); 
\node[below] at (4.5,2.4) {$-\varepsilon H_{\ell_3}$};

\draw[dashed] (-0.5,3.4) -- (6.5,3.4);
\node[left] at (-0.5,3.4) {$t_3$};

\fill[gray!90] (3,3.3) rectangle (4,3.5);
\node[above] at (3.5,3.5) {$-\varepsilon H_{\ell_4}$};

\end{tikzpicture}
\caption{Illustration of a continuous-time path-integral configuration $[\ell]$, introduced in \cref{eq:rhoth}, for a nearest-neighbor Hamiltonian in one spatial dimension. The configuration shown contains $N_\ell=4$ bond operators $(-\varepsilon H_{\ell_i})$ at times $t_i$ ($i=1,2,3,4$). Each operator is accompanied by a factor $\varepsilon$, representing the continuous-time measure $dt$, which cancels naturally during updates enforcing detailed balance. The full configuration $[\ell,s]$ is obtained by inserting complete sets of spin states, as explained in \cref{eq:rhoth-1}.}
\label{fig:CTconf}
\end{figure}

The thermal density matrix is defined by the matrix element
\begin{align}
\rho_{s_f,s_i}
&= \langle s_f| e^{-\beta H} |s_i\rangle
= \langle s_f| \underbrace{e^{-\varepsilon H} \cdots e^{-\varepsilon H}}_{L_t\ \text{terms}} |s_i\rangle ,
\end{align}
where the standard Trotter decomposition is introduced with $L_t$ time slices and temporal lattice spacing $\varepsilon$, such that $\beta = \varepsilon L_t$. As we explain in the next section, we can take the limit $\varepsilon \to 0$ and $L_t \to \infty$ in our method. In this limit we can expand each short-time exponential to leading order as
\begin{align}
e^{-\varepsilon H}
\simeq
\left(1 - \varepsilon \sum_\ell H_\ell\right) e^{-\varepsilon E_0}
+ \mathcal{O}(\varepsilon^2).
\end{align}
Substituting this into the Trotterized expression gives
\begin{align}
\rho_{s_f,s_i}
= e^{-\beta E_0}
\sum_{[\ell]}
\langle s_f|
(-\varepsilon H_{\ell_{N_\ell}}) \cdots
(-\varepsilon H_{\ell_2})(-\varepsilon H_{\ell_1})
|s_i\rangle ,
\label{eq:rhoth}
\end{align}
where $[\ell]$ denotes a continuous-time configuration consisting of an ordered sequence of $N_\ell$ bond operators inserted at imaginary times $t_1 < t_2 < \cdots < t_{N_\ell}$. We refer to $[\ell]$ as the link configuration. An illustration of such a configuration for a one-dimensional lattice system with nearest-neighbor interactions is shown in \cref{fig:CTconf}. The \ac{TFIM} studied in this work features link configurations of this type, whereas the \ac{SSM} includes bond operators involving both nearest-neighbor and diagonal sites.

The path integral is formulated in a complete basis of the local Hilbert space, defined by the expression
\begin{align}
I = \sum_{s} |s\rangle\langle s|.
\label{eq:MCbasis}
\end{align}
We will refer to $s$ in the above expression as ``spins,'' although they may represent any discrete set of quantum numbers.  
Inserting \cref{eq:MCbasis} between successive bond operators, we obtain the full path-integral expression
\begin{align}
\rho_{s_f,s_i}
&= e^{-\beta E_0}
\sum_{[\ell,s]}
\langle s_f|
(-\varepsilon H_{\ell_{N_\ell}})|s_{N_\ell-1}\rangle
\langle s_{N_\ell-1}| \ldots
\nonumber\\
&\quad \ldots
|s_{t_2}\rangle
\langle s_{t_2}|(-\varepsilon H_{\ell_2})|s_{t_1}\rangle
\langle s_{t_1}|(-\varepsilon H_{\ell_1})|s_i\rangle ,
\nonumber\\
&= e^{-\beta E_0} \sum_{[\ell,s]} W([\ell,s]),
\label{eq:rhoth-1}
\end{align}
where the sum now extends over both the link configurations $[\ell]$ and the intermediate spin states $[s]$, subject to the fixed boundary conditions $s_i$ and $s_f$. The combined configuration $[\ell,s]$ carries a weight $W([\ell,s])$, which is given by the product of local matrix elements of the form $\langle s_t | (-\varepsilon H_\ell) | s_{t-1} \rangle$. If all local matrix elements are positive, $W([\ell,s])$ is positive, and one can directly design a Monte Carlo algorithm to sample configurations $[\ell,s]$ with this weight. With a suitable choice of $E_0$, this positivity can be ensured for diagonal elements.  

However, off-diagonal elements may still be negative, rendering $W([\ell,s])$ sign-indefinite. To address this, we define a modified Hamiltonian $H_\ell^b$ by
\begin{align}
(-\varepsilon H_\ell^b) = \big| -\varepsilon H_\ell \big| ,
\end{align}
so that all matrix elements are nonnegative. Using this definition, we sample configurations according to the probability distribution
\begin{align}
P([\ell,s]) = \rho^b_{s_f,s_i} / Z^b ,
\end{align}
where
\begin{align}
\rho^b_{s_f,s_i}
&= e^{-\beta E_0}
\langle s_f|
(-\varepsilon H^b_{\ell_k}) \cdots
(-\varepsilon H^b_{\ell_2})(-\varepsilon H^b_{\ell_1})
|s_i\rangle
\nonumber\\
&= e^{-\beta E_0}
\sum_{[\ell,s]} W^b([\ell,s]),
\label{eq:rhothb}
\end{align}
and the normalization
\begin{align}
Z^b = \sum_{s_i,s_f} \rho^b_{s_f,s_i}
\end{align}
serves as the sampling partition function.

The effect of the original signs is restored through a sign factor ${\cal S}([\ell,s])$, defined as the product of the local signs of the matrix elements,
\begin{align}
W([\ell,s]) = {\cal S}([\ell,s])\, W^b([\ell,s]).
\end{align}
The thermal density matrix can then be expressed as
\begin{align}
\rho_{s_f,s_i}
= Z^b \Big\langle {\cal S}([\ell,s]) \Big\rangle ,
\label{eq:rhomc}
\end{align}
where the expectation value is evaluated as a Monte Carlo average over configurations $[\ell,s]$ sampled with weight $W^b([\ell,s])$ and fixed boundary states $s_i$ and $s_f$.

\section{The Monte Carlo Algorithm}
\label{sec3}

In this section, we summarize the two Monte Carlo updates used to generate the path-integral configurations $[\ell,s]$ discussed above. We refer to these as (i) the link update and (ii) the spin update. We conclude the section by explaining how the sampled configurations are used to compute the $Z$ and $E$ matrices introduced in \cref{eq:Zmat,eq:Emat} as observables.

\subsection{Link update}

The link update modifies the space–time locations of the link operators in the configuration $[\ell]$. Following the idea of the stochastic series expansion (SSE) algorithm \cite{Sandvik1999_SSE}, only diagonal link operators are inserted or removed during this update. A typical proposal (chosen with probability $1/2$) is either to add a new diagonal link operator or to remove an existing one. To add a link, a random spatial location (chosen uniformly among $V$ sites) and a random time slice at that location (chosen uniformly among $L_t$ slices) are selected, and one of the $d_\ell$ possible diagonal operators that can be inserted at that location is proposed. To remove a link, a random spatial location (chosen uniformly among $V$ sites) is selected, and one of the $n_\ell$ existing diagonal links at that location (across all time slices) is chosen uniformly and proposed for deletion. Each proposal is then accepted or rejected according to the Metropolis criterion, ensuring detailed balance.

To illustrate, let $W_i$ be the weight of the initial configuration and $W_f$ that of the proposed configuration obtained by adding a diagonal bond with matrix element $\varepsilon w = \langle s_{t_k}|(-\varepsilon H_{\ell_k}^b)|s_{t_k-1}\rangle$. This implies $W_f = W_i (\varepsilon w)$. If the number of diagonal links at the chosen location is $n_\ell$, it changes to $n_\ell + 1$ after insertion. The forward and reverse proposal probabilities (excluding the Metropolis acceptance factor) are then proportional to
\begin{align}
q_{i\to f} &= \tfrac{1}{2}\,\tfrac{1}{V}\,\tfrac{1}{L_t}, &
q_{f\to i} &= \tfrac{1}{2}\,\tfrac{1}{V}\,\tfrac{1}{(n_\ell + 1)}.
\end{align}
Detailed balance gives the Metropolis acceptance probability
\begin{align}
A_{i\to f} = \min\!\left(1,\; \frac{q_{f\to i}\,W_f}{q_{i\to f}\,W_i}\right),
\end{align}
and similarly for the reverse move. Because $W_f/W_i = (\varepsilon w)$ and $q_{f\to i}/q_{i\to f} = L_t/(n_\ell+1)$, the acceptance probabilities depend only on $\beta = \varepsilon L_t$. This demonstrates that the algorithm possesses a smooth continuous-time limit, allowing us to impose $\varepsilon\rightarrow 0$ and $L_t\rightarrow \infty$.

In practice, when implementing the algorithm, we set $\varepsilon \approx 10^{-8}$ (or smaller) and use a large number of time slices, $L_t = \beta / \varepsilon$. In this discrete-time approach, if a link already exists at the chosen time slice when a new link is proposed for insertion, no additional link is added. However, such occurrences become vanishingly rare as $\varepsilon$ decreases further. It is also worth noting that the memory requirements do not scale with $L_t$, since only the locations of the $N_\ell$ bonds in the configuration $[\ell,s]$ need to be stored, and this number scales with the physical space–time volume $V\beta$.

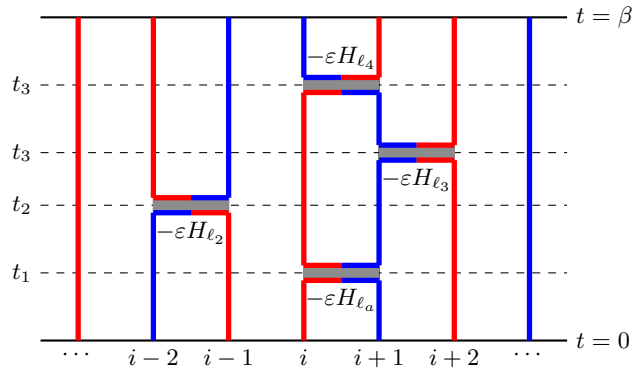
\begin{figure}[ht]
\centering
\begin{tikzpicture}

\draw[thick] (-0.5,0) -- (6.5,0) node[right] {$t=0$};
\draw[thick] (-0.5,4.3) -- (6.5,4.3) node[right] {$t=\beta$};

\foreach \x in {0,1,2,3,4,5,6}
    \draw[thin] (\x,0) -- (\x,4.3);

\node[below] at (0,0) {$\cdots$};
\node[below] at (1,0) {$i-2$};
\node[below] at (2,0) {$i-1$};
\node[below] at (3,0) {$i$};
\node[below] at (4,0) {$i+1$};
\node[below] at (5,0) {$i+2$};
\node[below] at (6,0) {$\cdots$};

\draw[dashed] (-0.5,0.9) -- (6.5,0.9);
\node[left] at (-0.5,0.9) {$t_1$};

\fill[gray!90] (3,0.8) rectangle (4,1.0);
\node[below] at (3.5,0.8) {$-\varepsilon H_{\ell_a}$};

\draw[dashed] (-0.5,1.8) -- (6.5,1.8);
\node[left] at (-0.5,1.8) {$t_2$};

\fill[gray!90] (1,1.7) rectangle (2,1.9); 
\node[below] at (1.5,1.7) {$-\varepsilon H_{\ell_2}$};

\draw[dashed] (-0.5,2.5) -- (6.5,2.5);
\node[left] at (-0.5,2.5) {$t_3$};

\fill[gray!90] (4,2.4) rectangle (5,2.6); 
\node[below] at (4.5,2.4) {$-\varepsilon H_{\ell_3}$};

\draw[dashed] (-0.5,3.4) -- (6.5,3.4);
\node[left] at (-0.5,3.4) {$t_3$};

\fill[gray!90] (3,3.3) rectangle (4,3.5);
\node[above] at (3.5,3.5) {$-\varepsilon H_{\ell_4}$};


\draw[line width=2pt,red] (0,0) -- (0,4.3);

\draw[line width=2pt,blue] (1,0) -- (1,1.7);
\draw[line width=2pt,blue] (1,1.7) -- (1.5,1.7);
\draw[line width=2pt,red] (1.5,1.7) -- (2,1.7);
\draw[line width=2pt,red] (2,0) -- (2,1.7);

\draw[line width=2pt,red] (1,1.9) -- (1,4.3);
\draw[line width=2pt,red] (1,1.9) -- (1.5,1.9);
\draw[line width=2pt,blue] (1.5,1.9) -- (2,1.9);
\draw[line width=2pt,blue] (2,1.9) -- (2,4.3);

\draw[line width=2pt,red] (3,0) -- (3,0.8);
\draw[line width=2pt,red] (3,0.8) -- (3.5,0.8);
\draw[line width=2pt,blue] (3.5,0.8) -- (4,0.8);
\draw[line width=2pt,blue] (4,0) -- (4,0.8);

\draw[line width=2pt,red] (3,3.3) -- (3,1.0);
\draw[line width=2pt,red] (3,1.0) -- (3.5,1.0);
\draw[line width=2pt,red] (3,3.3) -- (3.5,3.3);
\draw[line width=2pt,blue] (4,2.4) -- (4,1.0);
\draw[line width=2pt,blue] (4,2.6) -- (4,3.3);
\draw[line width=2pt,blue] (3.5,1.0) -- (4,1.0);
\draw[line width=2pt,blue] (3.5,3.3) -- (4,3.3);
\draw[line width=2pt,blue] (4,2.6) -- (4.5,2.6);
\draw[line width=2pt,blue] (4,2.4) -- (4.5,2.4);
\draw[line width=2pt,red] (4.5,2.6) -- (5,2.6);
\draw[line width=2pt,red] (4.5,2.4) -- (5,2.4);
\draw[line width=2pt,red] (5,2.6) -- (5,4.3);
\draw[line width=2pt,red] (5,2.4) -- (5,0);

\draw[line width=2pt,blue] (3,3.5) -- (3,4.3);
\draw[line width=2pt,blue] (3,3.5) -- (3.5,3.5);
\draw[line width=2pt,red] (3.5,3.5) -- (4,3.5);
\draw[line width=2pt,red] (4,3.5) -- (4,4.3);

\draw[line width=2pt,blue] (6,0) -- (6,4.3);

\end{tikzpicture}
\caption{The loop-cluster configuration  $[\ell,s]$ associated with the  continuous-time path-integral configuration $[\ell]$ shown in \cref{fig:CTconf}. The two colors represent opposite spins.}
\label{fig:Loopconf}
\end{figure}

\subsection{Spin update}

The spin update keeps the space–time locations of the link operators $(-\varepsilon H_\ell)$ fixed and updates the spin configuration $[s]$ in a manner consistent with the matrix elements of these operators. Several efficient methods now exist to implement changes to $[s]$ while maintaining detailed balance, including cluster updates \cite{Evertz:2000rk}, worm updates \cite{PhysRevLett.87.160601}, and loop updates \cite{PhysRevB.82.024407}.

In some cases, spin configurations can be reorganized into correlated clusters. For example, in the \ac{SSM}, nonzero matrix elements of the bond Hamiltonian always connect opposite spins on neighboring sites in the bra and ket states. This allows for the well known cluster representation of the spins \cite{PhysRevLett.70.875}. Imposing this restriction on the link configuration in \cref{fig:CTconf} generates the corresponding cluster configuration of spins, as illustrated in \cref{fig:Loopconf}. In our work, we employ such cluster representations, together with the meron-cluster approach \cite{Chandrasekharan:1999cm}, to partially alleviate some of the sign problems associated with the \ac{SSM}.

\subsection{Computation of
  \texorpdfstring{$Z$ and $E$}{Z and E} matrices}

Each new $[\ell,s]$ configuration is generated from an old $[\ell,s]$ configuration after a single sweep, which consists of a sequence of link updates followed by a sequence of spin updates. The numbers of link and spin updates in each sweep are chosen such that all bond operators have a reasonable probability of being updated during the sweep. The quoted results are obtained from statistics of sampled $[\ell,s]$ configurations generated over a few billion sweeps. We typically average observables over subsets of these configurations to construct thousands of data sets and then perform a bootstrap analysis over these averaged data sets to estimate the mean values and statistical errors.

Our observables are the projected thermal density matrix $Z$ and the projected thermal energy matrix $E$, defined in \cref{eq:Zmat,eq:Emat}. These quantities can be estimated as Monte Carlo averages using the wavefunctions $\psi_a(s) = \langle s|\psi_a\rangle$ of the projection states. One obtains
\begin{align}
\frac{Z_{ac}}{Z^b} &= \big\langle \mathcal{S}([\ell,s])\,\psi_a(s_f)\,\psi_c(s_i) \big\rangle_{W^b},
\label{eq:zmatmc_revised}\\
\frac{\widetilde{E}_{ac}}{Z^b} &= -\frac{1}{\beta}\big\langle N_\ell\, \mathcal{S}([\ell,s])\,\psi_a(s_f)\,\psi_c(s_i) \big\rangle_{W^b},
\label{eq:ematmc_revised}
\end{align}
where $\langle\cdot\rangle_{W^b}$ denotes an average over configurations sampled with weight $W^b([\ell,s])$. The sign factor $\mathcal{S}([\ell,s])$, introduced earlier in \cref{sec2}, corrects for the modified sampling weight. It is straightforward to verify that
\begin{align}
E Z^{-1} = \widetilde{E} Z^{-1} + E_0,
\end{align}
and that the common normalization factor $Z^b$ cancels in this calculation.

\section{Choice of the Projection Subspace}
\label{sec4}

The choice of the projection subspace is a crucial step in the successful application of our method. This subspace determines not only the projected low-energy spectrum of $H$, but also the projected thermal energies $e_a(\beta)$ and the rate at which these thermal energies converge to the low-energy spectrum. We now analyze this rate of convergence quantitatively for sufficiently large values of $\beta$, where this question becomes meaningful. As we will see, the convergence is exponential in $\beta$~\cite{Blossier:2009kd}.

Let us first define $\projS$ as the projection operator that maps the full Hilbert space onto the projection subspace. The projected thermal density matrix, in operator form, is then given by
\[
\hat{Z} = \projS e^{-\beta H} \projS,
\]
whose matrix elements in the chosen projection states were specified in \cref{eq:Zmat}. In general, several energy eigenstates $|\alpha\rangle$ of $H$ with energies ${\cal E}_\alpha$ will have nonzero overlap with the projection subspace. We will refer to these as the {\it projected energies} of $H$.
Each of these projected energies is associated with the {\it projected vector} $|p_\alpha\rangle = \projS|\alpha\rangle$ obtained by projecting $|\alpha\rangle$ onto the projection subspace. It is then straightforward to verify that
\begin{align}
\hat{Z} = \sum_\alpha |p_\alpha\rangle e^{-\beta {\cal E}_\alpha} \langle p_\alpha|,
\label{eq:ProjZmat}
\end{align}
where the sum is restricted to those eigenstates of $H$ that have a nonzero projection onto the projection subspace.

Note that the vectors $|p_\alpha\rangle$ are not necessarily orthogonal or even linearly independent. In fact, only $\dproj$ of them can be linearly independent. The central idea behind our method is that, for sufficiently large $\beta$, we can decompose \cref{eq:ProjZmat} into two terms:
\begin{align}
\hat{Z} &= \sum_{a=1}^\dproj |p_{\alpha_a}\rangle e^{-\beta {\cal E}_{\alpha_a}}\langle p_{\alpha_a}| 
+ \sum_\gamma |p_{\gamma}\rangle e^{-\beta {\cal E}_{\gamma}}\langle p_\gamma|.
\label{eq:projZmat1}
\end{align}
The first term contains a specific set of $\dproj$ linearly independent projected vectors, $\{|p_{\alpha_1}\rangle, \dots, |p_{\alpha_\dproj}\rangle\}$, which we refer to as the {\it projected low-energy vectors}. The second term arises from the remaining projected vectors that are linearly dependent on those in the first term. The vectors in the first term are associated with the lowest possible ordered energies ${\cal E}_{\alpha_1} \leq {\cal E}_{\alpha_2} \leq \dots \leq {\cal E}_{\alpha_\dproj}$ and define the {\it projected low-energy spectrum} introduced in \cref{sec1}, which we aim to compute using our method. The projected thermal energies $e_a(\beta)$ converge to ${\cal E}_{\alpha_a}$ in the limit $\beta \rightarrow \infty$. As we will see below, this term acts as a correction that governs the rate of convergence of $e_a(\beta)$.

We begin by outlining a systematic procedure to identify the projected low-energy vectors that contribute to the first term in \cref{eq:projZmat1}. We first construct an ordered list of all the vectors in \cref{eq:ProjZmat} according to their energies in ascending order. Within a degenerate subspace, the order can be chosen arbitrarily as long as the original eigenvectors of $H$ in the degenerate subspace are chosen carefully. We will come back to this point later. The first vector $|p_{\alpha_1}\rangle$ is identified as the lowest-energy vector on this ordered list. The next higher vector is identified as $|p_{\alpha_2}\rangle$ only if it is linearly independent of $|p_{\alpha_1}\rangle$; otherwise, we skip to the next higher-energy vector until an independent one is found. This process is repeated iteratively: $|p_{\alpha_k}\rangle$ is identified as the next independent vector following the first $k-1$ already identified vectors $\{|p_{\alpha_1}\rangle, |p_{\alpha_2}\rangle, \ldots, |p_{\alpha_{k-1}}\rangle\}$. The search terminates once $|p_{\alpha_\dproj}\rangle$ has been identified. All remaining vectors that are skipped or not encountered during this process contribute to the second term in \cref{eq:projZmat1}.

Each term in the degenerate subspace with a fixed low energy ${\cal E}_{\alpha_i}$ in the first term carries a suppression factor $e^{-\beta {\cal E}_{\alpha_i}}$. This subspace can also receive contributions from the second term, but they are always accompanied by a suppression factor $e^{-\beta {\cal E}_{\gamma}}$, where even the smallest ${\cal E}_{\gamma}$ is strictly greater than ${\cal E}_{\alpha_i}$. To ensure that ${\cal E}_{\gamma} > {\cal E}_{\alpha_i}$, the original eigenvectors of $H$ in the degenerate subspace of energy ${\cal E}_{\alpha_i}$ must be chosen such that any projected vector from that subspace appearing in the second term of \cref{eq:projZmat1} has no overlap with any projected vector in the degenerate subspace of the first term. This is always possible.

With this insight, one can use \cref{eq:projZmat1} to show that, in the $\beta \rightarrow \infty$ limit, the first term dominates. Keeping only this term, we find that $e_a(\beta \rightarrow \infty) = {\cal E}_{\alpha_a}$. The same expression can be used to estimate the rate of convergence, which is exponential, with corrections of the form $e^{-\beta \Delta_a}$, where the energy scale $\Delta_a$ is determined by the details of the second term in \cref{eq:projZmat1}. This rate can be as slow as $\Delta_a = {\cal E}_{\alpha_{a+1}} - {\cal E}_{\alpha_a}$ or as fast as $\Delta_a = {\cal E}_{\gamma} - {\cal E}_{\alpha_a}$, where ${\cal E}_{\gamma} > {\cal E}_{\alpha_\dproj}$.

An ideal projection subspace would be one constructed using the exact eigenstates of $H$. In that case, the second term in \cref{eq:projZmat1} vanishes, and the thermal energies satisfy $e_{\alpha_i}(\beta) = {\cal E}_{\alpha_i}$ even for arbitrarily small $\beta$, as discussed previously. This was used recently to compute Fermi energies of free fermions using a Monte Carlo that typically suffers from a severe sign problem \cite{Chandrasekharan:2024iao}. The next best scenario arises when perturbation theory provides a good approximation and the projection subspace of the free theory is used. In this situation, we obtain the best-case convergence with $\Delta_a = {\cal E}_{\gamma} - {\cal E}_{\alpha_\dproj}$~\cite{Blossier:2009kd}, and our method provides a non-perturbative approach that takes into account higher order corrections easily. Symmetries can also assist in constructing an effective projection subspace, since imposing symmetry constraints automatically restricts the energy eigenstates that have overlap with the chosen subspace.

In practical calculations, additional challenges arise when extracting the projected thermal energies $e_{\alpha_i}(\beta)$. These difficulties arise because the matrix elements of $Z/Z^b$ and $\widetilde{E}/Z^b$ generally become small as the system size $L$ or the inverse temperature $\beta$ increases. In particular, the matrix $Z/Z^b$ may become non-invertible. Even when $Z/Z^b$ remains invertible, it can contain very small eigenvalues that become negative due to Monte Carlo errors. This is unphysical and can make the projected thermal energies complex.
We will refer to these instabilities which result from signal to noise issues, as manifestations of the ``sign problem" inherent in the method. Thus, obtaining $e_a(\beta)$ with controlled uncertainties requires careful analysis.

In this work, we employ the following procedure to compute $e_a(\beta)$. We first evaluate the diagonal elements $Z_{aa}/Z^b$, which must be nonzero and positive with small relative errors. Failure to achieve this indicates the need for higher Monte Carlo statistics; if this is not feasible, the corresponding projection state $|\psi_a\rangle$ is removed from the analysis, thereby reducing the dimension of the projection subspace. After ensuring that all diagonal elements are well determined, we compute the eigenvalues of the matrix $Z/Z^b$. These eigenvalues must again be positive and nonzero within small relative errors. If this condition is not met, it implies that the Monte Carlo method is encountering a signal to noise issue, akin to the sign problem. In this situation, high-energy states above some energy threshold are thermally suppressed, and the Monte Carlo sampling becomes insensitive to their presence. However, without these high-energy states, there may no longer exist $\dproj$ independent projected vectors $|p_{\alpha_i}\rangle$ introduced earlier. In other words, the projected low-energy spectrum then contains fewer than $\dproj$ eigenvalues. Consequently, not all $\dproj$ projected effective thermal energies $e_a(\beta)$ remain meaningful. This issue manifests itself through vanishing eigenvalues of $Z$. In such cases, we discard the eigenvectors of $Z/Z^b$ whose eigenvalues are zero or negative within uncertainties and project both $Z/Z^b$ and $E/Z^b$ onto the remaining eigenvector subspace of $Z/Z^b$. The projected effective thermal energies are then computed from these reduced matrices. Because of this procedure, we cannot always extract all $\dproj$ projected effective thermal energies, but the subset of energies that can be determined in this way is numerically stable and reproducible within errors.

We end this section by discussing a special one-dimensional projection subspace that arises when the sign problem can be solved. In this case, there exists a state $|\psi_a\rangle$ whose diagonal matrix element satisfies $Z_{aa}/Z^b = 1$ for all values of $L$ and $\beta$. The corresponding one-dimensional projection subspace, $\projS_{(a)}$, spanned by $|\psi_a\rangle$, can then be used to compute the projected effective thermal energy as
\begin{align}
e_a(\beta) = E_{aa}(\beta)/Z^b = E_{aa}(\beta)/Z_{aa}.
\end{align}
The associated projected low-energy level—namely, the lowest energy eigenvalue whose eigenstate has a nonzero overlap with $|\psi_a\rangle$—can then be determined with high accuracy by evaluating $e_a(\beta)$ at successively larger values of $\beta$, since the sign problem is absent. This idea is very similar to the projector Monte Carlo techniques discussed previously \cite{PhysRevLett.95.207203,PhysRevB.82.024407}.

In the above example, although the sign problem is absent for the one-dimensional projection subspace constructed from $|\psi_a\rangle$, it is likely to reappear once the subspace is enlarged by adding additional states $|\psi_c\rangle$. This is because the matrix $Z/Z^b$ may develop small eigenvalues as $\beta$ increases, which can be difficult to compute accurately using Monte Carlo methods.

\section{The Transverse Field Ising Model}
\label{sec5}

As a first application of our method, we extract the lowest four energy levels at zero momentum, ${\cal E}_0$, ${\cal E}_1$, ${\cal E}_2$, and ${\cal E}_3$, of the \ac{TFIM} at criticality in the presence of a small longitudinal field $h$. Zamolodchikov conjectured that the magnetic perturbation drives the critical Ising model towards an $E_8$ massive quantum field theory in the infrared (IR), which contains eight stable particles with precisely predicted mass ratios. Below the two particle continua, there are three masses $m_i = {\cal E}_i - {\cal E}_0$ from which we can create two universal ratios,
\begin{align}
\frac{m_2}{m_1} &= 2\cos(\pi/5) \approx 1.618, \\
\frac{m_3}{m_1} &= 2\cos(\pi/30) \approx 1.989,
\label{eq:E8pred}
\end{align}
In the lattice theory, this prediction holds only in the scaling limit, where the lattice size $L$ is large and the magnetic perturbation $h$ is small, and the thermal perturbation is zero (tuned by the transverse field). Our goal here is to demonstrate that the \ac{PDMS} method can be used to confirm \cref{eq:E8pred}. To the best of our knowledge, previous Monte Carlo attempts to extract this ratio were not successful \cite{Destri:1992aa}, although it has been confirmed by other numerical and analytic approaches\cite{Henkel:1989jn,Fonseca:2006au,PhysRevB.83.020407,albert2025,Jha:2024jan,litvinov2025meson}. For recent 

\subsection{The Hamiltonian}

The Hamiltonian of the \ac{TFIM} at criticality in the presence of a longitudinal magnetic field $h$ is given by
\begin{align}
H = -\sum_{i=0}^{L-1} \Big( \sigma_{i}^{z}\sigma^{z}_{i+1} + \sigma_{i}^{x} + h\,\sigma_{i}^{z} \Big),
\label{eq:TFIMH}
\end{align}
where $i$ labels the sites on a periodic chain of length $L$, and $\sigma_{i}^{\alpha}$ are Pauli matrices that act on the two-dimensional Hilbert space of the $i^{\mathrm{th}}$ site.

We construct the continuous-time path integral, as discussed in \cref{sec2}, using the basis states $|+\rangle$ and $|-\rangle$, which are eigenstates of the $\sigma^x_i$ operator on each lattice site with eigenvalues $+1$ and $-1$, respectively. We refer to these as the $X$-basis. For this purpose, we express $H$ in the form described in \cref{eq:mcH} and identify
\begin{align}
H_\ell = \Big(
&- \sigma_i^z\sigma_{i+1}^z 
- \frac{h}{2} (\sigma_i^z + \sigma_{i+1}^z) \nonumber \\
& - \frac{1}{2}\big[(1+\sigma_i^x) + (1+\sigma_{i+1}^x)\big]
\Big),
\label{eq:TFIMHl}
\end{align}
and $E_0 = L$, which makes all diagonal matrix elements in the X-basis negative semi-definite.

Each $H_\ell$ is a nearest-neighbor bond operator that connects sites $i$ and $i+1$ such that all matrix elements of $(-\varepsilon H_\ell)$ in the X-basis are positive. The link updates and spin updates, as discussed in \cref{sec4} are straightforward. When $h \neq 0$, the Ising symmetry is broken, but spin updates can still be constructed as explained in \cite{PhysRevE.67.046701}.

\begin{figure}[ht]
\centering
\def\Scale{0.6}
\begin{tikzpicture}[scale=\Scale,transform shape]

  \def\N{12}
  \def\R{2.2}
  \def\dwrad{0.06}
  \def\dwroff{0.12}

  \def\stemw{0.08}
  \def\stemh{0.40}
  \def\headw{0.24}
  \def\headh{0.28}
  \pgfmathsetmacro{\downshift}{\stemh-\headh}

  \pgfmathsetmacro{\phA}{90 - (4.5-1)*360/\N}
  \pgfmathsetmacro{\phB}{90 - (8.5-1)*360/\N}
  \pgfmathsetmacro{\rotbase}{(180 - (\phA+\phB))/2}
  \pgfmathsetmacro{\rot}{\rotbase - 180}

  \newcommand{\uparrowglyph}{%
    \filldraw[fill=blue,draw=black,line join=round]
      (-\stemw,-\stemh) -- (-\stemw,0) -- (-\headw,0) --
      (0,\headh) -- (\headw,0) -- (\stemw,0) --
      (\stemw,-\stemh) -- cycle;
  }
  \newcommand{\downarrowglyph}{%
    \begin{scope}[yshift={-\downshift cm}]
      \filldraw[fill=red,draw=black,line join=round]
        (-\stemw,\stemh) -- (-\stemw,0) -- (-\headw,0) --
        (0,-\headh) -- (\headw,0) -- (\stemw,0) --
        (\stemw,\stemh) -- cycle;
    \end{scope}
  }

  \foreach \i [evaluate=\i as \th using {\rot + 90-(\i-1)*360/\N},
               evaluate=\th as \xx using {\R*cos(\th)},
               evaluate=\th as \yy using {\R*sin(\th)}] in {1,2,3,4}{
    \begin{scope}[shift={(\xx,\yy)}] \uparrowglyph \end{scope}
  }
  \foreach \i [evaluate=\i as \th using {\rot + 90-(\i-1)*360/\N},
               evaluate=\th as \xx using {\R*cos(\th)},
               evaluate=\th as \yy using {\R*sin(\th)}] in {5,6,7,8}{
    \begin{scope}[shift={(\xx,\yy)}] \downarrowglyph \end{scope}
  }
  \foreach \i [evaluate=\i as \th using {\rot + 90-(\i-1)*360/\N},
               evaluate=\th as \xx using {\R*cos(\th)},
               evaluate=\th as \yy using {\R*sin(\th)}] in {9,10,11,12}{
    \begin{scope}[shift={(\xx,\yy)}] \uparrowglyph \end{scope}
  }

  \foreach \k [evaluate=\k as \ph using {\rot + 90-(\k-1)*360/\N},
               evaluate=\ph as \cx using {(\R-\dwroff)*cos(\ph)},
               evaluate=\ph as \cy using {(\R-\dwroff)*sin(\ph)}] in {4.5,8.5}{
    \draw[draw=black,fill=green] (\cx,\cy) circle[radius=\dwrad cm];
  }

\end{tikzpicture}
\caption{A 2-domain wall state of length $\ell=4$ for a chain of length $L=12$. States like these make up our projection subspace.}
\label{fig:domainwall}
\end{figure}
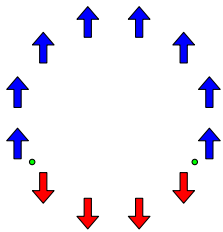

\begin{figure*}
\centering
\includegraphics[width=0.3\linewidth]{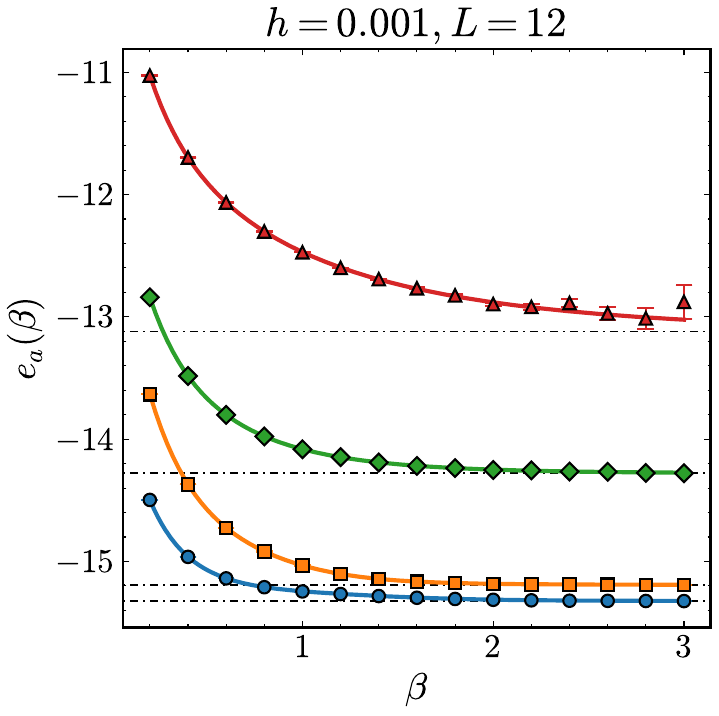}
\includegraphics[width=0.3\linewidth]{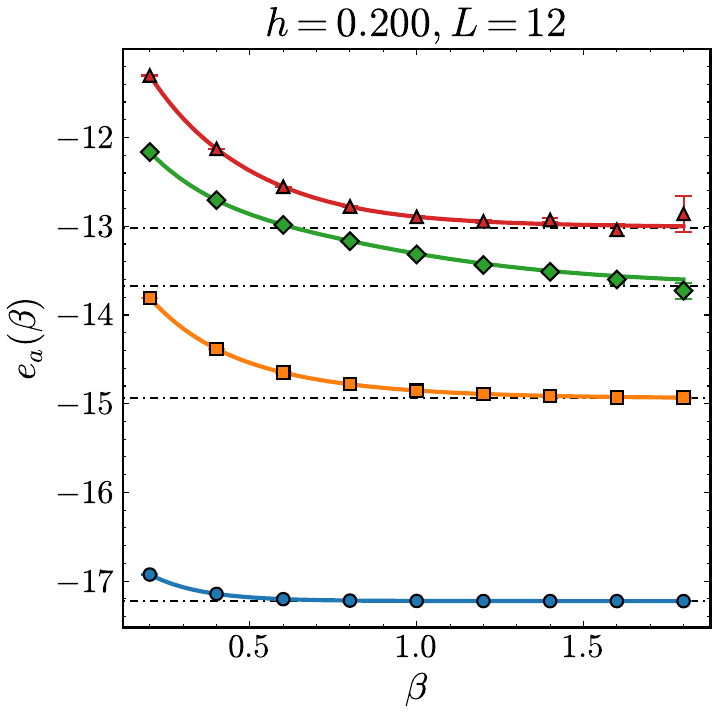}
\includegraphics[width=0.3\linewidth]{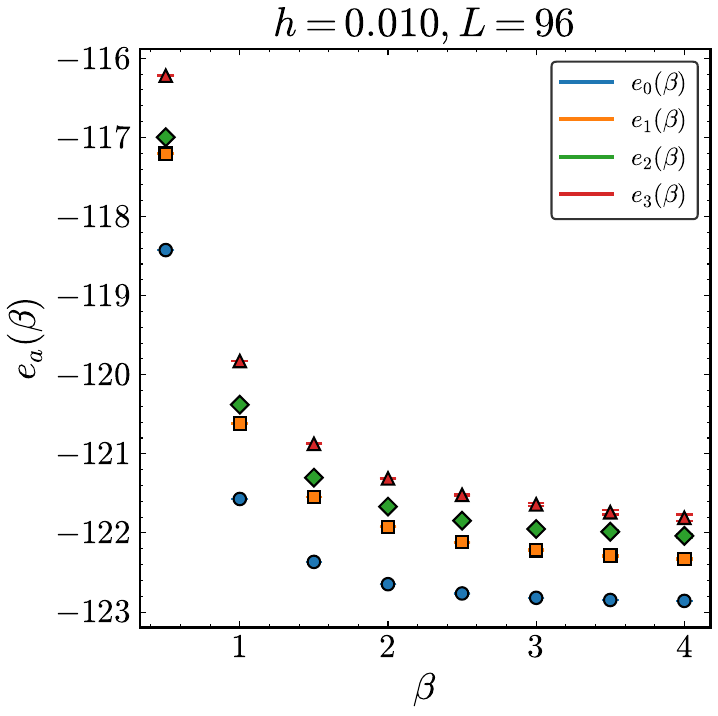}
\caption{The lowest four projected effective thermal energies obtained using the \ac{PDMS} method for 
$(L = 12,\, h = 0.001)$ (left), $(L = 12,\, h = 0.2)$ (center), and $(L = 96,\, h = 0.01)$ (right). 
The solid lines passing through the Monte Carlo data in the left and center panels represent exact diagonalization results. 
The dashed lines in these panels indicate the projected low-energy spectrum to which $e_a(\beta)$ is expected to converge in the $\beta \rightarrow \infty$ limit. 
The fourth level in the left and center panels begins to exhibit larger errors around $\beta \approx 3$, which we attribute to its proximity to the two-particle threshold, as discussed in the text. The right panel shows that the convergence of the energy levels is relatively slow due to the small gaps in the energy spectrum. Because the energies are extensive quantities, they scale with $L$, whereas the energy gaps do not. Consequently, obtaining accurate estimates of the gaps requires high statistics.}
\label{fig:L12comp}
\end{figure*}

\subsection{Projection Subspace}

We now discuss the choice of the projection subspace $\projS$ used to compute the lowest four eigenvalues of $H$ at zero momentum for various values of $h$ and $L$. Because the longitudinal field breaks the $\mathbb{Z}_{2}$ symmetry, momentum is the only remaining lattice symmetry that can be used to guide the construction of $\projS$. The emergent $E_8$ masses are therefore purely dynamical.

Since our path integral is formulated in the $X$-basis, simple states in this basis generally have exponentially small overlap with the low-energy states. In contrast, projection states chosen in the $Z$-basis—whose elements are eigenstates of $\sigma_i^z$—can have a much larger overlap. For example, consider the projection state
\begin{align}
|\psi_1\rangle = 2^{-L/2}\, |\uparrow\uparrow\uparrow\ldots\uparrow\rangle.
\label{eq:projTFIM1}
\end{align}
Using the relations $|\uparrow\rangle = (|+\rangle + |-\rangle)/\sqrt{2}$ and $|\downarrow\rangle = (|+\rangle - |-\rangle)/\sqrt{2}$, one can verify that $Z_{11}/Z^b = 1$, independent of $L$ and $\beta$. As discussed at the end of \cref{sec4}, this implies that the sign problem is absent for this particular one-dimensional projection subspace.

Using exact diagonalization studies on the $L = 12$ lattice, as described in \cref{app1}, we added additional states to the projection subspace. One of these is the $\mathbb{Z}_2$-flipped partner of \cref{eq:projTFIM1}, given by
\begin{align}
|\psi_2\rangle = 2^{-L/2}\, |\downarrow\downarrow\downarrow\ldots\downarrow\rangle.
\label{eq:projTFIM2}
\end{align}
While this state seems to have a good overlap with third excited state on small lattices, it causes instabilities on larger lattices. We drop it in our analysis for $L\geq 64$.
We also found that states in which a contiguous block of $\ell$ neighboring spins is flipped have a strong overlap with the low-energy eigenstates. States of this form can be viewed as two–domain-wall configurations, as illustrated in \cref{fig:domainwall}. Translating these states and summing over all translations projects them onto the zero-momentum sector. We enumerate them as
\begin{align}
|\psi_{\ell+2}\rangle = \frac{2^{-L/2}}{\sqrt{L}} \sum_{r=1}^{L} T^{r} 
|\underbrace{\downarrow\downarrow\ldots\downarrow}_{\ell}\uparrow\uparrow\ldots\uparrow\rangle,
\label{eq:projTFIMl}
\end{align}
where $\ell = 1, 2, \ldots, L/2$. Here, $T$ denotes the translation operator that shifts all spins by one lattice site to the right, and the summation enforces projection to zero momentum. In this work, we restrict $\ell \leq L/2$, although higher values ($L/2 < \ell < L$) could in principle be used. Thus, the total dimension of the projection subspace employed here is $\dproj = L/2 + 2$.

\begin{table*}[t]
\centering
\renewcommand{\arraystretch}{1.2}
\begin{tabular}{|c|c|c|c|c|c|c|}
\TopRule
$L$ & $\beta$ & $h$ & $e_0(\beta)$ & $e_1(\beta)$ & $e_2(\beta)$ & $e_3(\beta)$ \\
\MidRule
\multirow{18}{*}{16} & \multirow{2}{*}{2} & 0.071493 & $-21.2396(7)$ & $-19.883(15)$ & $-19.348(19)$ & $-18.68(7)$ \\
 &  & 0.085771 & $-21.4240(9)$ & $-19.913(14)$ & $-19.21(4)$ & $-18.63(11)$ \\
\cline{2-7}
 & \multirow{6}{*}{2.5} & 0.002427 & $-20.40307(23)$ & $-20.28470(25)$ & $-19.5943(15)$ & $-18.524(11)$ \\
 &  & 0.003640 & $-20.41362(21)$ & $-20.27502(20)$ & $-19.5933(17)$ & $-18.553(15)$ \\
 &  & 0.003980 & $-20.41644(23)$ & $-20.27194(27)$ & $-19.5924(15)$ & $-18.530(11)$ \\
 &  & 0.004320 & $-20.41960(26)$ & $-20.26887(27)$ & $-19.5983(16)$ & $-18.551(11)$ \\
 &  & 0.007393 & $-20.45014(24)$ & $-20.23879(21)$ & $-19.5977(17)$ & $-18.558(12)$ \\
 &  & 0.009085 & $-20.46848(21)$ & $-20.22112(24)$ & $-19.6002(15)$ & $-18.559(15)$ \\
\cline{2-7}
 & \multirow{9}{*}{3} & 0.010778 & $-20.48825(24)$ & $-20.2173(4)$ & $-19.6076(28)$ & $-18.518(34)$ \\
 &  & 0.016556 & $-20.55444(18)$ & $-20.1596(6)$ & $-19.6125(26)$ & $-18.60(4)$ \\
 &  & 0.022333 & $-20.62319(20)$ & $-20.1021(8)$ & $-19.6143(28)$ & $-18.63(5)$ \\
 &  & 0.028111 & $-20.69247(22)$ & $-20.0492(9)$ & $-19.6134(25)$ & $-18.70(5)$ \\
 &  & 0.033889 & $-20.76298(19)$ & $-20.0042(16)$ & $-19.6039(30)$ & $-18.57(7)$ \\
 &  & 0.039667 & $-20.83495(20)$ & $-19.9668(20)$ & $-19.597(4)$ & $-18.70(8)$ \\
 &  & 0.045444 & $-20.90689(15)$ & $-19.930(4)$ & $-19.5755(35)$ & $-18.56(7)$ \\
 &  & 0.051222 & $-20.97981(15)$ & $-19.913(4)$ & $-19.553(4)$ & $-18.78(9)$ \\
 &  & 0.057000 & $-21.05325(20)$ & $-19.906(5)$ & $-19.514(6)$ & $-18.54(13)$ \\
\cline{2-7}
 & \multirow{1}{*}{3.5} & 0.001213 & $-20.40623(23)$ & $-20.30395(15)$ & $-19.6180(21)$ & $-18.68(4)$ \\
\MidRule\MidRule
\multirow{4}{*}{32} & \multirow{4}{*}{4} & 0.015000 & $-41.0715(4)$ & $-40.4724(31)$ & $-40.3061(17)$ & $-39.836(16)$ \\
 &  & 0.020000 & $-41.18964(33)$ & $-40.495(4)$ & $-40.210(4)$ & $-39.905(30)$ \\
 &  & 0.026650 & $-41.3490(4)$ & $-40.549(10)$ & $-40.096(22)$ & $-39.91(5)$ \\
 &  & 0.033300 & $-41.5117(4)$ & $-40.610(9)$ & $-40.06(9)$ & $-39.77(9)$ \\
\MidRule\MidRule
\multirow{3}{*}{64} & \multirow{3}{*}{3.5} & 0.011500 & $-82.0233(21)$ & $-81.451(9)$ & $-81.059(18)$ & $-80.862(22)$ \\
 &  & 0.015000 & $-82.1653(20)$ & $-81.518(11)$ & $-81.121(27)$ & $-80.69(8)$ \\
 &  & 0.018500 & $-82.3128(16)$ & $-81.615(14)$ & $-81.17(4)$ & $-80.05(28)$ \\
\MidRule\MidRule
\multirow{3}{*}{96} & \multirow{3}{*}{4.5} & 0.010000 & $-122.8625(8)$ & $-122.339(4)$ & $-122.040(10)$ & $-121.798(27)$ \\
 &  & 0.012000 & $-123.0026(8)$ & $-122.434(6)$ & $-122.130(18)$ & $-121.870(35)$ \\
 &  & 0.013500 & $-123.1048(7)$ & $-122.494(6)$ & $-122.144(21)$ & $-121.82(4)$ \\
\BotRule
\end{tabular}
\caption{Parameters $(h,L,\beta)$ used to obtain the mass ratio $m_2/m_1$ as a function of $\mu$ displayed in \cref{fig:scaling_ftn}. At these parameters we can use the \ac{PDMS} method to reliably extract the mass ratio for $\mu\in[0.446,9.663]$ which covers both the Ising \ac{CFT} and the $E_8$ \ac{QFT}.}
\label{tab:ranges}
\end{table*}

\begin{table}[t]
\centering
\renewcommand{\arraystretch}{1.15}
\begin{tabular}{|r|c|c|c|}
\hline
\multicolumn{1}{|c|}{$L$} & \multicolumn{1}{c|}{$\beta$-range} & \multicolumn{1}{c|}{$h_{\parallel}$-range} & \multicolumn{1}{c|}{$\mu$-range} \\
\hline
16 & 2.0-3.0 & 0.001-0.0858 & 0.446-4.318 \\
\hline
32 & 4.0 & 0.015-0.033 & 3.407-5.213 \\
\hline
64 & 3.5 & 0.012-0.019  & 5.194-7.621 \\
\hline
96 & 4.0 & 0.01-0.0135 & 8.234-9.663 \\
\hline
\end{tabular}
\caption{Parameters $(h_{\parallel},L,\beta)$ used to obtain the mass ratio $r_{1}$ as a function of $\mu$ displayed in \cref{fig:scaling_ftn}. Theses parameters were chosen for computational tractability of the \ac{PDMS} method.}
\label{tab:TFIMres}
\end{table}

\begin{figure*}
\centering
\includegraphics[width=0.7\linewidth]{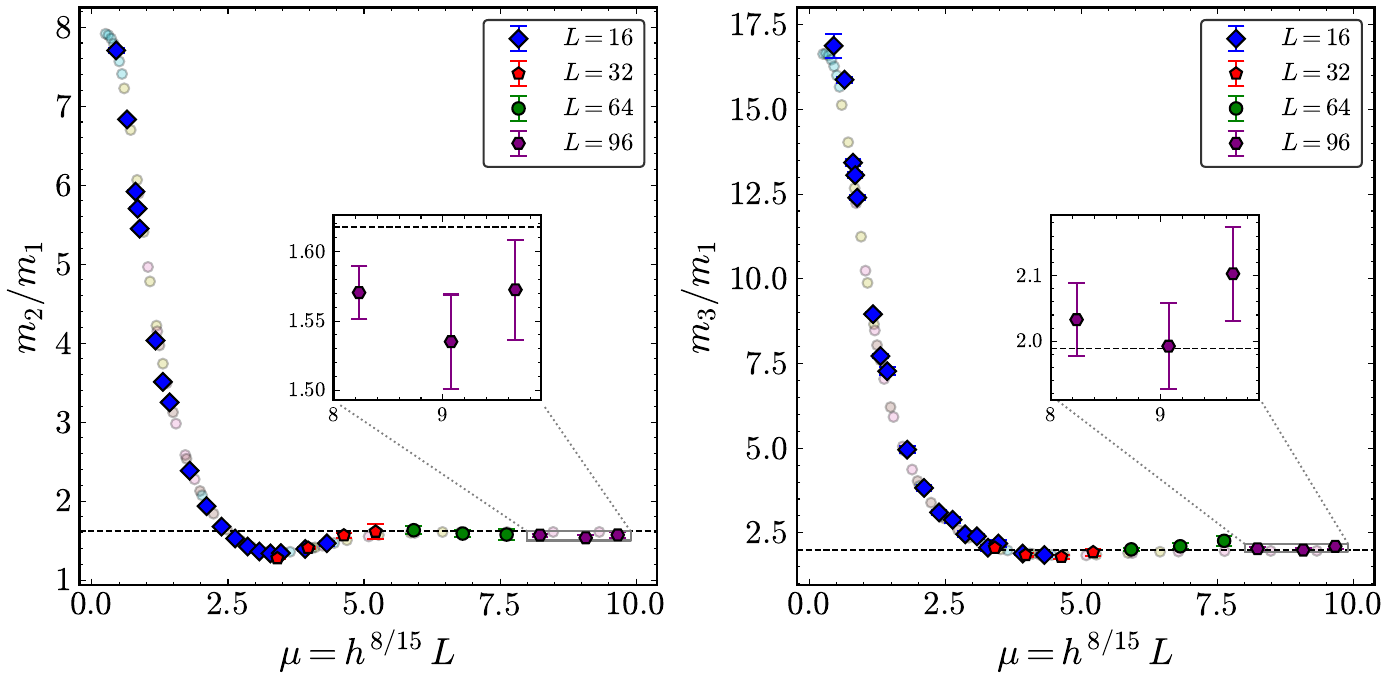}
\caption{Monte Carlo estimate of the mass ratio $m_{2}/m_{1}$ (left) and $m_{3}/m_{1}$ (right) as a function of $\mu$. The dashed horizontal line shows the expected $E_8$ \ac{QFT} result of $m_{2}/m_{1}=1.618$ and $m_{3}/m_{1}=1.989$. The inset shows our estimate at some of the largest values of $\mu$ we have computed. The light dots give estimates from exact diagonalization results at various small values of $L$ and $h$.}
\label{fig:scaling_ftn}
\end{figure*}

\subsection{Projected Thermal Energies}

We have used the \ac{PDMS} method to compute the lowest four masses for various lattice sizes $L$, inverse temperatures $\beta$, and longitudinal field values $h$. While the full projection subspace $\projS$ is used to extract these quantities, care must be taken in removing zero modes in the $Z/Z^b$ matrix, as discussed in \cref{sec4}. Additional details of this analysis are provided in \cref{app1}.

To confirm the reliability of the method, we first verified that it reproduces exact diagonalization results. In \cref{fig:L12comp}, we compare Monte Carlo data with exact diagonalization results for $L = 12$ at two values of $h$. The Monte Carlo data accurately reproduce the exact results for the first three energy levels. However, the fourth level, $e_3(\beta)$, begins to exhibit larger error bars around $\beta \approx 3$. We attribute this behavior to the fact that this level corresponds to a massive particle with $m_3 \approx 1.989\, m_1$ (see \cref{eq:E8pred}), which lies close to the two-particle threshold at $2m_1$. Consequently, it receives contributions from several two–$m_1$ particle states that naturally appear at energies above $2m_1$. This makes it difficult to extract $m_3$ reliably, beyond noting that its value is close to $2m_1$.

After confirming that the \ac{PDMS} method accurately reproduces the lowest four energy levels at $L = 12$, we extended our calculations to a range of lattice sizes $L$, temperatures $\beta$, and longitudinal fields $h$ to test the prediction of \cref{eq:E8pred}. The resulting values are summarized in \cref{tab:TFIMres}. We note that the raw energy levels can be computed reasonably accurately, although the relative errors of thermal energy gaps $e_i(\beta)-e_0(\beta)$ deteriorate as the energies increase.

\subsection{Extraction of \texorpdfstring{$E_8$}{E8} Mass Ratio}

It is well known that the two-dimensional massive $E_8$ \ac{QFT} emerges from a relevant magnetic perturbation of the Ising \ac{CFT}. In the critical \ac{TFIM}, introducing a small longitudinal field $h$ generates a correlation length that scales as $\xi \sim h^{-8/15}$. If this \ac{CFT} is probed on a finite lattice with $L$ sites, a simple scaling hypothesis suggests that all observables in the theory become functions of a single dimensionless variable,
\begin{align}
\mu = L\, h^{8/15}.    
\end{align}
One immediate consequence is that the mass ratios $m_i/m_1$ should depend only on $\mu$. In the limit $\mu = 0$, the system approaches the Ising \ac{CFT}, where, for example, $m_2/m_1 = 8$. In the opposite limit $\mu \to \infty$, the theory begins to describe the $E_8$ \ac{QFT}, where $m_2/m_1 = 1.618$.

In a lattice calculation, any finite value of $L$ introduces corrections that break conformal invariance, so the above statements hold strictly only in the simultaneous limits $h \to 0$ and $L \to \infty$, with $\mu$ held fixed. In practice, however, the scaling hypothesis is already observed to work well even for moderate system sizes. For example, as shown in \cref{fig:L12comp}, when $h = 0.001$ (corresponding to $\mu = 0.301$), we obtain $m_2/m_1 = 7.888(37)$, close to the Ising \ac{CFT} prediction of $8$. Conversely, when $h = 0.2$ (corresponding to $\mu = 5.086$), we find $m_2/m_1 = 1.526(40)$, consistent with the $E_8$ \ac{QFT} prediction of $1.618$.

To compute the mass ratios in the $E_8$ \ac{QFT}, we must focus on the limit $\mu \rightarrow \infty$ with $h \rightarrow 0$ and $L \rightarrow \infty$. Reaching this regime numerically requires identifying a set of parameters $(h, L, \beta)$ for which the projected thermal energies $e_a(\beta)$ have converged to the projected low-energy spectrum within small statistical errors. 

The main difficulty at large $L$ with fixed $\mu$ is that the energy gaps decrease while each individual energy scales with $L$. Consequently, accurately determining the gaps requires substantial statistics, and larger values of $\beta$ are needed for convergence. This challenge is compounded by the fact that the physical Hilbert space grows exponentially with $L$, whereas the projection subspace grows only linearly, leading to a reduced overlap between any energy eigenstate of $H$ and the projection subspace. 

Conversely, using smaller $L$ and larger $h$ increases the energy gaps. However, in this regime, the matrix $Z/Z^b$ can become nearly singular at large $\beta$, since some of its eigenvalues scale as $e^{-\beta m_i}$ and may vanish within Monte Carlo errors. This is effectively a manifestation of the sign problem, which renders $Z/Z^b$ non-invertible. 

Despite these challenges, we have identified a regime of parameters where it is still possible to reliably extract the ratio $m_2/m_1$. The corresponding parameter ranges are listed in \cref{tab:ranges}, and the results for the lowest four projected thermal energies in this regime are summarized in \cref{tab:TFIMres}.

Our final results for the ratios $m_2/m_1$ and $m_3/m_1$ as functions of $\mu = L\, h^{8/15}$, obtained from the \ac{PDMS}, are shown in \cref{fig:scaling_ftn}. We find that the ratios saturate to the expected $E_8$ value for $\mu \gtrsim 6$, within expected errors. We have verified that our results are consistent with the known universal scaling curve for these ratios as a function of $\mu$  \cite{Henkel:1989jn}.

\section{The Shastry--Sutherland Model}
\label{sec6}

As a second application, in this section, we investigate whether the \ac{PDMS} method can be applied to study the highly frustrated Shastry--Sutherland model (SSM) at various coupling strengths. Since the model is known to suffer from a severe sign problem, our first goal here is modest: we aim to reproduce the values of the projected thermal energies $e_a(\beta)$ for small lattice sizes and small $\beta$, which can also be computed using exact diagonalization. These results can be found in \cref{sec6-D,sec6-E}.

Our second goal is to demonstrate that our method readily extends to larger lattices beyond the reach of exact diagonalization methods, as long as $\beta$ is kept small. We further argue that even relatively small values of $\beta$ can sometimes yield results that are close to the zero-temperature limit, provided a suitable projection subspace is chosen. To this end, we focus on two different regimes of the SSM on an $L=8$ lattice, where we can compare results from our method with those obtained from other reliable approaches in earlier work. In \cref{sec6-F}, we show that we can accurately compute the lowest three energy levels of the two-dimensional Heisenberg model on the square lattice (the unfrustrated limit of the \ac{SSM}) using data at small values of $\beta$. In \cref{sec6-G}, we examine the first excited energy level of the \ac{SSM} at $(J=1.0, J'=0.5)$, which is accessible in perturbation theory. We find that the energy gap between this level and the ground state at $L=4$ is already in excellent agreement with perturbation theory, suggesting that larger lattice sizes are unnecessary. This conclusion is supported by our calculations at $L=8$, which yield results consistent within statistical errors with those obtained at $L=4$. Calculations at $L=8$ are noisier, and reaching larger values of $\beta$ becomes increasingly challenging.

In all these studies of the SSM model, we consider two simple projection subspaces that distinguish between two symmetry sectors of the model and examine the challenges involved in extracting the projected low-energy spectrum associated with these subspaces.

\subsection{The Hamiltonian}
\label{sec6-A}

\begin{figure}[ht]
\centering
\includegraphics[width=0.45\textwidth]{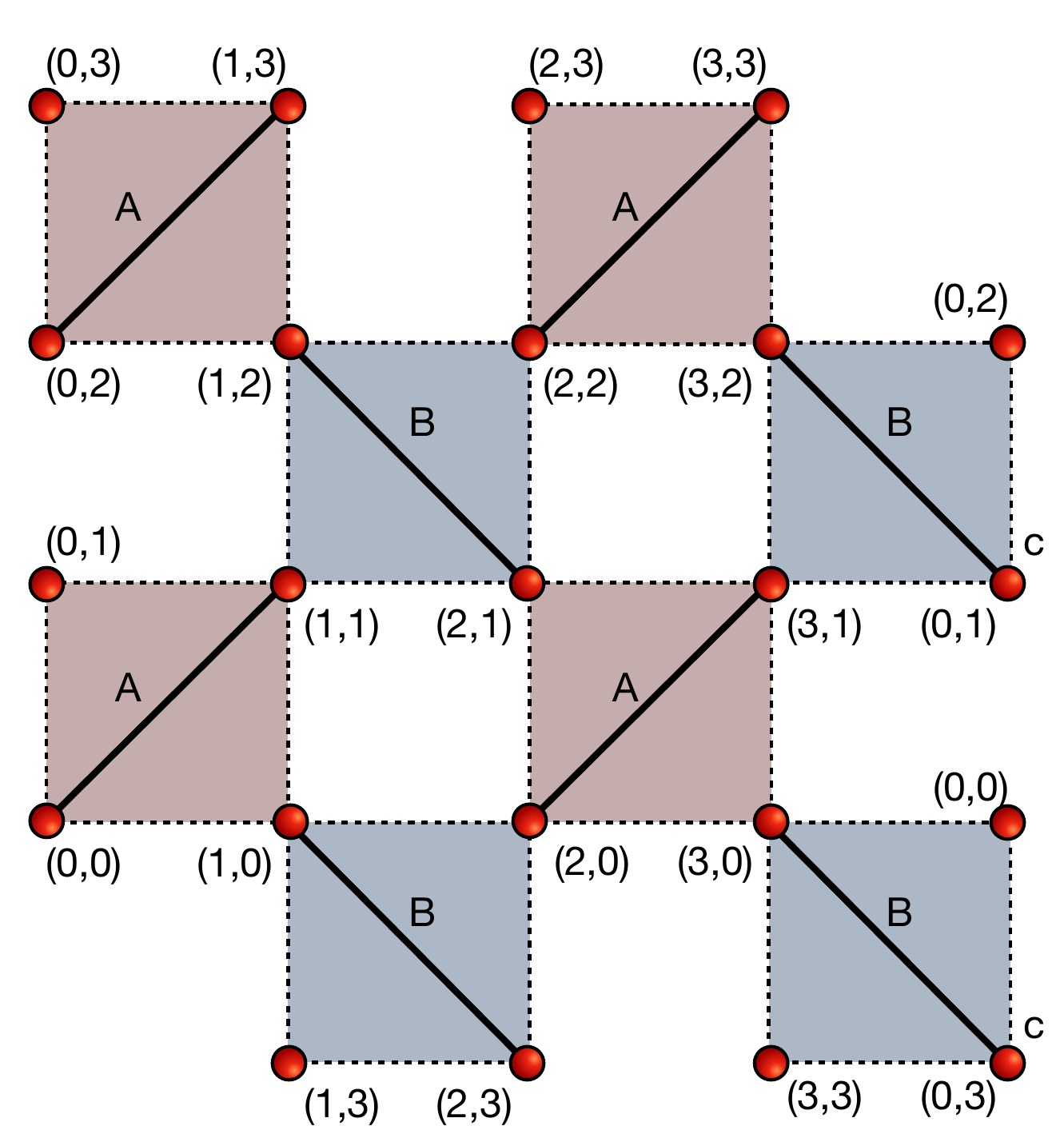}
\caption{Illustration of the Shastry–Sutherland lattice on a $4\times4$ periodic system. Thin dashed black lines indicate the $\nn$ bonds with interaction weight $J^\prime$ bonds, while thick solid lines represent the $\nnn$ bonds with interaction weight $J$. The coordinates of each site are also shown for convenience.}
\label{fig:SSMlattice}
\end{figure}

The Hamiltonian of the SSM is given by
\begin{align}
H = J' \sum_{\nn} \bs_i \cdot \bs_j + J \sum_{\nnn} \bs_i \cdot \bs_j,
\label{eq:ssm}
\end{align}
where $i,j$ label sites on an $L \times L$ periodic square lattice. The quantum spin-$\tfrac{1}{2}$ operators $\bs_i = (s_i^x, s_i^y, s_i^z)$ associated with site $i$ interact antiferromagnetically with operators $\bs_j$ on neighboring sites $j$ through two types of bonds, denoted by $\nn$ for the nearest-neighbor bonds and $\nnn$ for a specific set of diagonal bonds. An illustration of these bonds on a $4 \times 4$ periodic lattice is shown in \cref{fig:SSMlattice}. Further details of our convention for labeling the lattice and the bonds can be found in \cref{app2}.

We argued in \cref{sec2} that it is important to rewrite $H$ in the form
\begin{align}
H = \sum_{\ell} H_\ell + E_0,
\end{align}
to construct the path integral. For the \ac{SSM}, $\ell$ refers to interactions on both $\nn$ and $\nnn$ bonds. We define
\begin{align}
H_\ell = J_{ij} \left(\bs_i \cdot \bs_j - \tfrac{1}{4}\right),
\label{eq:hlssm}
\end{align}
where $J_{ij}$ equals either $J$ or $J'$, depending on whether $(i,j)$ corresponds to an $\nn$ or $\nnn$ bond, respectively. It is straightforward to verify that
\begin{align}
E_0 = \sum_{i,j} J_{ij} = \left(\tfrac{J'}{2} + \tfrac{J}{8}\right) L^2,
\label{eq:e0ssm}
\end{align}
where $E_0$ is chosen such that all diagonal matrix elements of $H_\ell$ are either zero or negative in the chosen basis. 

Using \cref{eq:hlssm}, we can construct the path integral and design a Monte Carlo algorithm as discussed in \cref{sec2,sec3}. We employ the conventional $s_i^z$ eigenbasis, labeled by $|\ua\rangle$ and $|\da\rangle$, although other types of dimer bases could be more effective in mitigating sign problems \cite{PhysRevB.57.R3197}. To partially alleviate the sign problem in the $s_i^z$ basis, we work in the spin-cluster representation and include the entropy factor that arises from summing over an entire class of spin configurations during the bond updates. This helps in reducing the fluctuations while computing the $Z$ and $E$ matrices using \cref{eq:zmatmc_revised,eq:ematmc_revised}.

\subsection{Projection Subspaces}
\label{sec6-B}

Symmetries often provide valuable guidance in constructing suitable projection subspaces. Since the SSM is invariant under spin rotations, it is useful to define projection subspaces that isolate a single spin sector $s$ with a specific magnetic quantum number $m$. However, in this work, we consider a simpler conserved quantity associated with the operator
\begin{align}
\hat{Q} = \prod_i (2 s_i^x),
\end{align}
whose eigenvalues are $\pm 1$. Because $[H, \hat{Q}] = 0$, all energy eigenstates can be classified according to their $\hat{Q}$ quantum number. Based on $\hat{Q}$, we define two projection subspaces, $\projS_+$ and $\projS_-$, which distinguish energy eigenstates with eigenvalues $Q = +1$ and $Q = -1$, respectively. These subspaces are constructed as direct-product states of single-site and two-site (dimer) basis states. Details of how these local states are defined are provided in \cref{app2}.

Our $\projS_+$ subspace is four-dimensional and constructed as the span of four simple projection states defined as
\begin{align}
|\psi_1\rangle &= \bigotimes_{i \in \mathrm{even}} (|\ua\rangle + |\da\rangle)_i  
\bigotimes_{j \in \mathrm{odd}} (|\da\rangle - |\ua\rangle)_j, \nonumber \\
|\psi_2\rangle &= \bigotimes_{\nnn} |\psi_t\rangle_{\nnn}, \nonumber \\
|\psi_3\rangle &= \bigotimes_{\substack{\nn \\ x_i\,\mathrm{even},\, x_j = x_i + 1 \\ x_i\,\mathrm{odd},\, x_j = x_i - 1}} |\psi_s\rangle_{\nn}, \nonumber\\
|\psi_4\rangle &= \bigotimes_{\substack{\nn \\ x_i\,\mathrm{even},\, x_j = x_i + 1 \\ x_i\,\mathrm{odd},\, x_j = x_i - 1}} |\psi_t\rangle_{\nn},
\label{eq:psibasis}
\end{align}
where the dimer product states $|\psi_s\rangle$ and $|\psi_t\rangle$ are defined in \cref{eq:singlet,eq:triplet}. These states are illustrated for $L = 2$ in \cref{fig:2x2projst} and for $L = 4$ in  \cref{fig:4x4projst}. It is straightforward to verify that all of these states are eigenstates of the $\hat{Q}$ operator with eigenvalue $+1$.

It is also straightforward to verify that another $Q = +1$ state, constructed as the product of singlet dimers on the $\nnn$ bonds,
\begin{align}
|\psi_0\rangle = \bigotimes_{\substack{\nnn,\, x_i\,\mathrm{even}}} 
|\psi_s\rangle_\nnn,
\label{eq:dimerstate}
\end{align}
is an eigenstate of $H$ with eigenvalue 
\begin{align}
{\cal E}_S = -\frac{3}{8} J L^2,
\end{align}
for all values of $(J, J')$. One of the central results of Ref.~\cite{ShastrySutherland1981} is that, for $J' \leq J/2$, $|\psi_0\rangle$ is also the ground state of $H$.

When $J' = 0$, since the $\nnn$ dimers do not interact, all the excited states can be constructed exactly. The first excited state, which has an energy gap of order $J$, can be obtained by placing all $\nnn$ dimers in the spin-singlet state except for one, which is placed in the triplet state $|\psi_t\rangle$ defined in \cref{eq:triplet}. The state in which the triplet is connected to the origin $(x_i, y_i) = (0,0)$ is given by
\begin{align}
|\phi_1\rangle = |\psi_t\rangle_{\substack{\nnn \\ x_i\,\mathrm{even}}} \bigotimes_{{\nnn}' \neq \nnn} |\psi_s\rangle_{{\nnn}'}.
\label{eq:phibasis}
\end{align}
The other states in the subspace, $|\phi_a\rangle$ ($a = 2, 3, \ldots, L^2/2$), are obtained by translating the triplet to other $\nnn$ bonds. Based on our discussion in \cref{sec4}, an ideal projection subspace for studying the excited spectrum in the perturbative regime of small $J'$ would be this $L^2/2$-dimensional degenerate subspace of first excited states at $J' = 0$. This subspace naturally has $Q = -1$, as can be verified, and forms our $\projS_-$ subspace. The various states in this subspace for $L=2$ and $L=4$ are pictorially illustrated in the appendix in \cref{fig:2x2projst,fig:4x4projst}.

For small $J'$, the states in $\projS_-$ mix only at high order in perturbation theory. This suggests that $\langle \phi_a | e^{-\beta H} | \phi_b \rangle \propto \delta_{ab}$ would be a good approximation, at least for small values of $J'$. Interestingly, we find that this behavior persists even at larger values of $J'$, partly due to lattice symmetries and partly due to dynamical effects. While at small $J'$ the $\projS_-$ subspace has a large overlap with the first excited state, at larger values of $J'$ it mixes only weakly with lower-energy states, indicating that it is an optimal subspace for exploring the projected low-energy spectrum only for small values of $J'/J$.

\subsection{Choice of Couplings}

We focus on four values of the couplings $(J, J')$, motivated by the physics of the SSM, and compute the projected thermal energies for small lattice sizes $L$ and inverse temperatures $\beta$. The goal is to explore the challenges involved in extracting the projected low-energy spectrum associated with $\projS_+$ and $\projS_-$.  

Among the four couplings, the first case we study is $(J = 0, J' = 1)$, where the \ac{SSM} reduces to the 2D-unfrustrated Heisenberg quantum antiferromagnet. Here, the low-energy physics is that of a spontaneously broken continuous symmetry with a dense spectrum of low-lying excitations. Our goal in this case is to examine how well the $\projS_+$ subspace reproduces the low-energy spectrum. The second coupling we consider is $(J = 1, J' = 0.5)$, where the \ac{SSM} lies in a gapped phase. Here, our main objective is to determine whether the energy gap can be extracted using the $\projS_-$ subspace. We then investigate the couplings $(J = 1, J' = 0.66)$ and $(J = 0, J' = 0.75)$, which lie close to two phase transitions predicted in the \ac{SSM}~\cite{Corboz2025} using both the projection subspaces to learn how they perform.

\begin{figure*}[t]
\centering
\includegraphics[width=\textwidth]{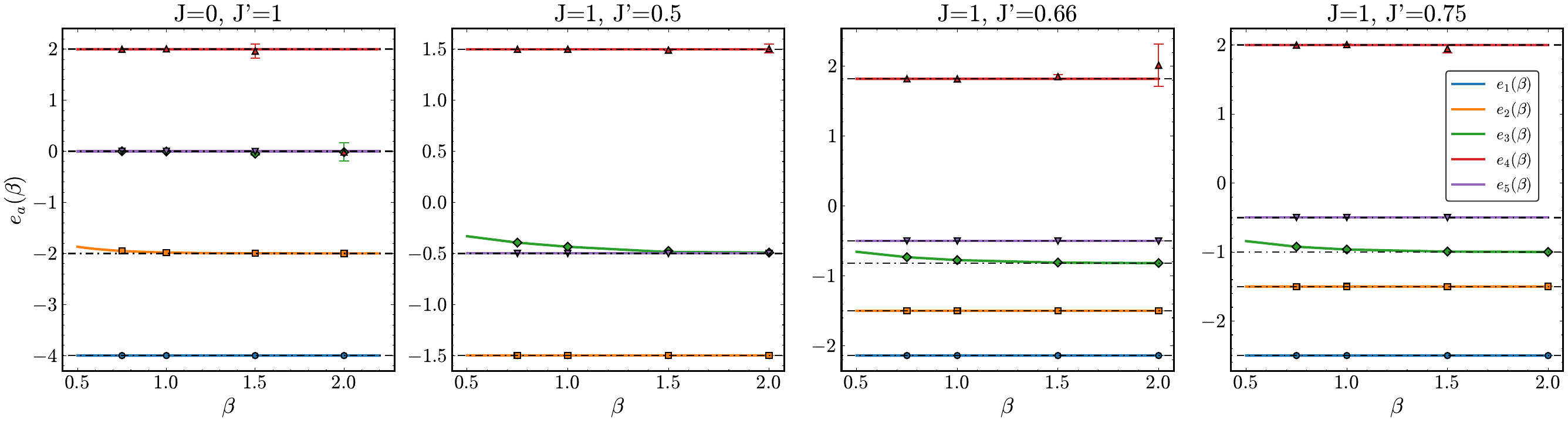}
\caption{The projected thermal energies extracted using the two projection subspaces $\projS_+$ and $\projS_-$ discussed for $L=2$. The two distinct energies in the $\projS_-$ subspace are degenerate and hence only one is shown. The Monte Carlo data was obtained only at $\beta=0.75,1,1.5,2$. The solid lines are exact diagonalization results. The projected low energy spectrum is shown as horizontal dashed lines and their quantitative values can be found in \cref{tab:L2projE}. While in general there are five distinct projected low energy levels, at $(J=0, J'=1)$ and $(J=1, J'=0.5)$ some levels become degenerate as can be seen from \cref{tab:L2projE}.}
\label{fig:L2plots}
\end{figure*}

\begin{table}[t]
\centering
\renewcommand{\arraystretch}{1}
\begin{tabular}{|M{0.3\linewidth}|M{0.3\linewidth}|M{0.3\linewidth}|}
\TopRule
\multirow{4}{*}{Coupling} & \multicolumn{2}{|c|}{Projected Low Energy Spectrum} \\
 \cline{2-3}
& $\projS_+$ & $\projS_-$ \\
\cline{2-3}
& ${\cal E}_1,{\cal E}_2,{\cal E}_3,{\cal E}_4$ & ${\cal E}_1={\cal E}_2$ \\
\MidRule
($J=0$, $J'=1$) & $E_1,E_2,E_3,E_5$ & $E_4(2)$\\
($J=1$, $J'=0.5$) &$E_1,E_1,E_2,E_3$ & $E_2(2)$ \\
($J=1$, $J'=0.66$) & $E_1,E_2,E_3,E_5$ & $E_4(2)$ \\
($J=1$, $J'=0.75$) & $E_1,E_2,E_3,E_5$ & $E_4(2)$ \\
\BotRule 
\end{tabular}
\caption{Projected low energy spectrum at various couplings at $L=2$. The energies $E_a$ are listed in \cref{tab:L2evals}. In the $\projS_-$ subspace the two energies are degenerate. The levels in the projected low energy spectrum are shown as dashed lines in \cref{fig:L2plots}.}
\label{tab:L2projE}
\end{table}

\subsection{Results at \texorpdfstring{$L = 2$}{L = 2}}
\label{sec6-D}

In this section, we present our results for the projected thermal energies $e_a(\beta)$ at $L = 2$ for both projection subspaces and examine how they converge to the projected low-energy spectrum. To identify this spectrum, it is first useful to distinguish the distinct energy levels of $H$. We denote these as $E_a$ ($a = 1, 2, \ldots$), ordered such that $E_1 \leq E_2 \leq E_3 \leq \ldots$. Each level is characterized by its spin representation $s$ and a degeneracy factor $g$ arising from exact lattice symmetries. The total degeneracy of level $E_a$ is therefore $g(2s + 1)$. If different spin representations exhibit additional accidental degeneracies, the corresponding energies are treated as distinct levels. Using these energies, together with information about the overlap of the associated energy eigenstates with the projection subspace, we can identify the projected low-energy spectrum ${\cal E}_1, {\cal E}_2, \ldots$, discussed in \cref{sec4}. This allows us to study the convergence of $e_a(\beta)$ to ${\cal E}_a$.  

For $L = 2$, all eigenstates of $H$ can be constructed as direct products of states on the two diagonal bonds, as discussed in \cref{app2}. The distinct ordered energies $E_a$ for the four chosen couplings are summarized in \cref{tab:L2evals}. Using the relations provided in \cref{app2} and following the procedure outlined in \cref{sec4}, we can identify the projected low-energy spectrum in the two different subspaces. These results are shown in \cref{tab:L2projE}. Note that since the dimension of the $\projS_+$ subspace is four, we identify four energies. Some of these energies are accidentally degenerate at certain couplings. However, although the dimension of the $\projS_-$ subspace is two, we display only one energy because both are identical due to lattice symmetry as explained in \cref{sec6-B}.  

A comparison between the projected thermal energies obtained from our Monte Carlo data and the exact results at all couplings is shown in \cref{fig:L2plots}. As the figure illustrates, these energies converge exponentially to the projected low-energy spectrum, as expected, even at small values of $\beta$. Hence, it seems the chosen projection subspaces are nearly ideal. We obtain all five distinct energy eigenvalues with high accuracy even at $\beta = 1.0$.

\begin{table}[t]
\centering
\renewcommand{\arraystretch}{1}
\begin{tabular}{|M{0.3\linewidth}|M{0.3\linewidth}|M{0.3\linewidth}|}
\TopRule
\multirow{5}{*}{Coupling} & \multicolumn{2}{|c|}{Projected Low Energy Spectrum} \\
 \cline{2-3}
& $\projS_+$ & $\projS_-$ \\
\cline{2-3}
& \multirow{2}{*}{${\cal E}_1,{\cal E}_2,{\cal E}_3,{\cal E}_4$} & ${\cal E}_1=\ldots={\cal E}_4$, \\
& & ${\cal E}_5=\ldots={\cal E}_8$ \\
\MidRule
($J=0$, $J'=1$) & $E_1,E_2,E_3,E_7$ & $E_4 (4),E_5(4)$\\
($J=1$, $J'=0.5$) &$E_1,E_4,E_6,E_{32}$ & $E_2(4),E_3(4)$ \\
($J=1$, $J'=0.66$) & $E_1,E_2,E_3,E_9$ & $E_5(4),E_7(4)$ \\
($J=1$, $J'=0.75$) & $E_1,E_2,E_3, E_8$ & $E_5(4),E_6(4)$ \\
\BotRule 
\end{tabular}
\caption{Projected Low Energy Spectrum at various couplings at $L=4$. The energies $E_a$ are listed \cref{tab:L4evals}. In the $\projS_-$ subspace the eight energies split into two distinct levels each of which contains four degenerate states. The levels in the projected low energy spectrum are shown as dashed lines in \cref{fig:L4plots}.}
\label{tab:L4projE}
\end{table}

\begin{figure*}[t]
\centering
\includegraphics[width=\textwidth]{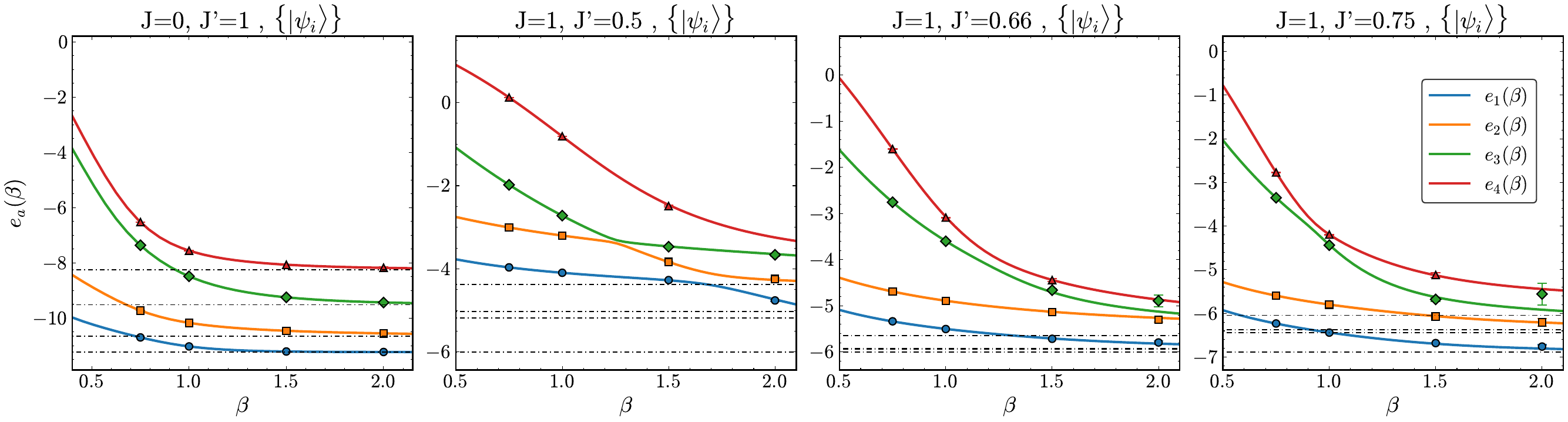}
\includegraphics[width=\textwidth]{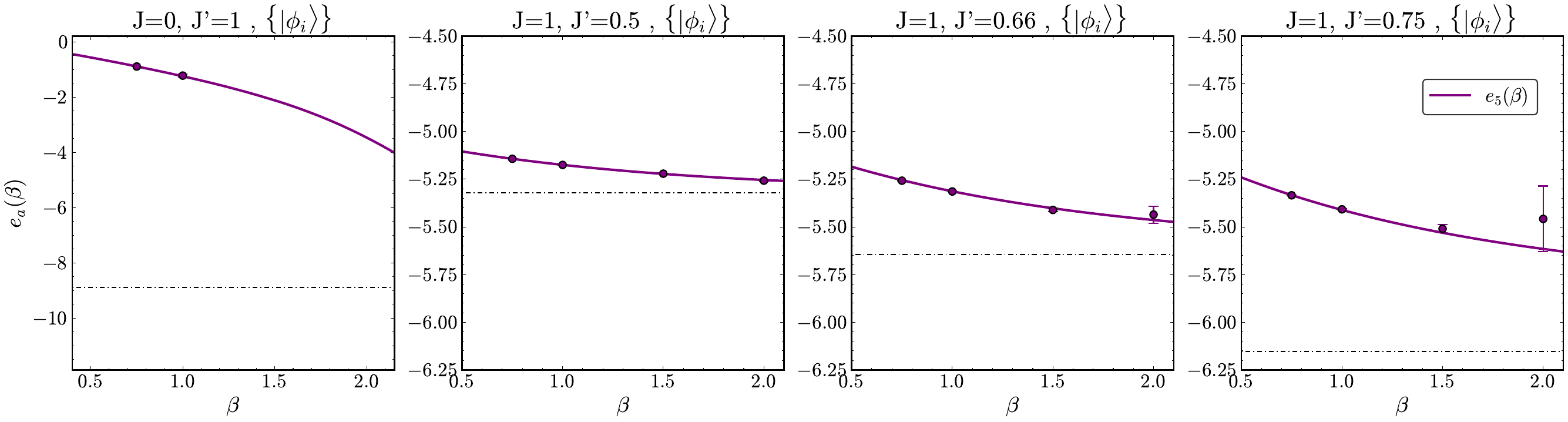}
\caption{Projected thermal energies $e_a(\beta)$ extracted using $\projS_+$ (top row) and $\projS_-$ (bottom row) for $L = 4$. While $\projS_-$ is eight-dimensional, the energies split into two four-dimensional degenerate subspaces with two distinct but nearly equal values. Hence, we show only one representative degenerate energy for the entire subspace. The Monte Carlo data were obtained at $\beta = 0.75, 1.0, 1.5, 2.0$. The solid lines represent exact diagonalization results, which consistently agree with the Monte Carlo data, demonstrating the validity of our \ac{PDMS} method. The projected low-energy spectrum is shown as horizontal dashed lines in each panel, and their numerical values can be found in \cref{tab:L4projE}. Missing data indicate the presence of a sign problem and the necessity to omit certain states during the analysis. The numerical values of the Monte Carlo data at $\beta = 2$ and the corresponding dashed-line energies are listed in \cref{tab:L4datacomp}.}
\label{fig:L4plots}
\end{figure*}

\begin{table*}[t]
\centering
\renewcommand{\arraystretch}{1.2}
\begin{tabular}{|M{0.1\linewidth}|M{0.1\linewidth}|M{0.1\linewidth}|M{0.1\linewidth}|M{0.1\linewidth}|M{0.1\linewidth}|M{0.1\linewidth}|M{0.1\linewidth}|M{0.1\linewidth}|}
\TopRule
& \multicolumn{2}{|M{0.2\linewidth}|}{$(J=0,J'=1)$} &
\multicolumn{2}{|M{0.2\linewidth}|}{$(J=1,J'=0.5)$} &
\multicolumn{2}{|M{0.2\linewidth}|}{$(J=1,J'=0.66)$} &
\multicolumn{2}{|M{0.2\linewidth}|}{$(J=1,J'=0.75)$} \\
\TopRule
Projection Subspace & Exact & Monte Carlo  &
Exact  & Monte Carlo  &
Exact  & Monte Carlo  &
Exact  & Monte Carlo  \\
& ($\beta\rightarrow \infty$) & ($\beta = 2$) &
($\beta\rightarrow \infty$) & ($\beta = 2$) &
($\beta\rightarrow \infty$) & ($\beta = 2$) &
($\beta\rightarrow \infty$) & ($\beta = 2$) \\
\MidRule
\multirow{4}{*}{$\projS_+$} & -11.22848... & -11.2269(2) &
-6.00000... & -4.76(2) &
-6.00000... & -5.83(4) &
-6.88440.... & -6.79(5)\\
& -10.64988... & -10.5524(2) &
-5.18260... & -4.26(3) &
-5.93309... & -5.33(6) &
-6.44305... & -6.34(6)\\
& -9.51768... & -9.436(3) &
-5.02962... & -3.68(4) &
-5.91680... & -5.1(1) &
-6.37022... & -5.9(3)\\
& -8.25282... & -8.17(2) &
-4.37361... & -3.1(2) &
-5.62894... & -4.7(2) &
-6.04382... & -4.7(4)\\
\MidRule
$\projS_-$ & -8.79437... & - &
-5.32028... &   &
-5.64562... & -5.44(5) &
-6.15238.... & -5.5(2)\\
\BotRule
\end{tabular}
\caption{Comparison of the projected thermal energies $e_a(\beta)$ at $\beta = 2$ with the projected low-energy spectrum ($e_a(\beta = \infty)$) at various couplings for $L = 4$. The first four rows correspond to $\projS_+$, while the last row lists the degenerate level in the eight-dimensional $\projS_-$ subspace. The missing value indicates the presence of a sign problem.}
\label{tab:L4datacomp}
\end{table*}

\subsection{Results at \texorpdfstring{$L = 4$}{L = 4}}
\label{sec6-E}

We now repeat the above analysis for $L = 4$. In this case, we use exact diagonalization to compute the eigenvalues and eigenvectors of $H$. In \cref{tab:L4evals}, we list the lowest 45 distinct energy levels $E_a$, along with their spin quantum numbers $s$ and representation degeneracies $g$, for the four coupling sets considered. From this table, we observe that the dimer state of \cref{eq:dimerstate}, with energy ${\cal E}_S = -6$, is the ground state at both $(J = 1, J' = 0.5)$ and $(J = 1, J' = 0.66)$. However, it becomes the $a = 10$ level at $(J = 1, J' = 0.75)$ and no longer appears among the lowest 45 distinct levels when $(J = 0, J' = 1)$.

The first and second excited states also evolve with the couplings. For $(J = 1, J' = 0.5)$, these two states are nearly degenerate triplets with energies $E_1 \approx E_2 \approx -5.320$, each triplet being fourfold degenerate. These states overlap with the eight-dimensional $\projS_-$ subspace. The overlap information for some of the low-energy states with $|\phi_1\rangle$ is shown in \cref{tab:L4ov-phi}. In fact, the overlaps with all eight projection states $|\phi_i\rangle$, $i = 1, \ldots, 8$, are very similar, and hence only one of them is displayed. As argued earlier, the energy splitting among the states that overlap with $\projS_-$ is expected to be small at small $J'$, and this behavior is indeed visible even at $J' = 0.5$. More accurately the eight degenerate first excited states at $J'=0$, split into two distinct four-fold degenerate levels given by $E_2$ and $E_3$ at $J'=0.5$. These then naturally form the projected low-energy spectrum of the $\projS_-$ subspace at $(J = 1, J' = 0.5)$. The energy gap between these states and the ground state is approximately $0.68$. In contrast, using the overlap information in \cref{tab:L4ov-psi}, we observe that the projected low-energy spectrum of the $\projS_+$ subspace is $E_1$, $E_4$, $E_6$, and $E_{32}$.

At $(J = 1, J' = 0.66)$, the lowest-energy excitations are no longer among the $\projS_-$ states. These triplets now correspond to $E_5$ and $E_7$, each forming a four-dimensional subspace, with splittings that remain small. These states constitute the projected low-energy spectrum of $\projS_-$. Several new $s = 0$ states appear below them. The first and second excited states are singlets separated by a very small gap of about $0.015$, with the second excited state being doubly degenerate, most likely due to a lattice symmetry. The energy gap between the first excited state and the ground state is even smaller, roughly $0.07$. Using \cref{tab:L4ov-psi}, we find that the projected low-energy spectrum of $\projS_+$ consists of $E_1$, $E_2$, one of the $E_3$ states, and $E_9$.

At $(J = 1, J' = 0.75)$, the ground state changes to a distinct singlet with energy around $-6.88$. The first excited state is another singlet, doubly degenerate due to lattice symmetry, separated by a gap of approximately $0.44$. The second excited state is a triplet lying about $0.07$ above the first excited state. Using information from \cref{tab:L4ov-psi}, we conclude that these four states (i.e., $E_1$, $E_2$, $E_3$, and $E_8$) form the projected low-energy spectrum of the $\projS_+$ subspace. From \cref{tab:L4ov-phi}, we observe that the states $E_5$ and $E_6$, both fourfold degenerate, form the projected low-energy spectrum of the $\projS_-$ subspace. Again, the energy splitting among them remains small, though somewhat larger than for smaller values of $J'$.

Finally, for $(J = 0, J' = 1)$, the ground state is a singlet, followed by a triplet and a quintuplet as the first and second excited states, respectively—consistent with expectations for an antiferromagnet. Using \cref{tab:L4ov-psi}, we find that the projected low-energy spectrum of the $\projS_+$ subspace is $E_1$, $E_2$, $E_3$, and $E_7$. From \cref{tab:L4ov-phi}, we see that the two fourfold-degenerate subspaces with energies $E_4$ and $E_5$ form the projected low-energy spectrum of the $\projS_-$ subspace. However, \cref{tab:L4ov-phi} also shows that the overlaps of these states—and even of some higher-energy states—are quite small. Hence, we do not expect the $\projS_-$ subspace to be a good choice at $(J = 0, J' = 1)$, and the corresponding projected thermal energies are expected to converge very slowly.

The discussion above regarding the projected low-energy spectrum of both $\projS_+$ and $\projS_-$ subspaces at various couplings is summarized in \cref{tab:L4projE}. As discussed in \cref{sec4}, the rate at which the projected thermal energies converge to the projected low-energy spectrum depends on the second term in \cref{eq:projZmat1}, which involves the states $|p_\gamma\rangle$ that capture the overlap of higher excited states with the projection subspaces. In a generic many-body quantum system, the energy spectrum becomes increasingly dense at higher energies—this is also the case for the \ac{SSM}. As seen in \cref{tab:L4ov-phi} and \cref{tab:L4ov-psi}, our projection subspaces have significant overlap with several higher-energy states. Thus, we do not expect them to provide an optimal basis in which the projected thermal energies $e_a(\beta)$ rapidly converge to the projected low-energy spectrum.

The convergence of the projected thermal energies is shown in \cref{fig:L4plots}. In these plots, the solid lines represent exact diagonalization results, while the data points with error bars correspond to our Monte Carlo estimates. In all cases, the Monte Carlo results agree with the exact values within statistical uncertainties, confirming the reliability of our \ac{PDMS} method. The horizontal dashed lines indicate the projected low-energy spectrum (also listed in \cref{tab:L4projE}); these are the levels that the solid curves approach as $\beta \to \infty$, a trend we have verified with exact calculations. Although our two projection subspaces are not optimal in general, they perform well in several cases. For instance, $\projS_+$ helps us compute the first four levels of the projected low energy spectrum with reasonable accuracy at $(J = 0, J' = 1)$, while $\projS_-$ performs well in computing the first excited states at $(J = 1, J' = 0.5)$. For $(J = 1, J' = 0.66)$, the lowest level in the $\projS_+$ projection subspace has converged to within about three percent, while the degenerate level in the $\projS_-$ subspace lies about five percent away and exhibits larger statistical errors. We see similar results for the case at $(J = 1, J' = 0.75)$; $\projS_+$ allows us to compute the ground-state energy to within one percent whereas the $\projS_-$ subspace remains far from convergence. A quantitative comparison between the projected low-energy spectrum and $e_a(\beta)$ at $\beta = 2$ is presented in \cref{tab:L4datacomp}.

\begin{table*}[t]
\centering
\renewcommand{\arraystretch}{1.6}
\begin{tabular}{|M{0.1\linewidth}|M{0.2\linewidth}|M{0.12\linewidth}|M{0.12\linewidth}|M{0.12\linewidth}|M{0.12\linewidth}|}
\TopRule
\multirow{2}{*}{$\beta$}
 & Projected subspace $\projS_{(1)}$ &
\multicolumn{4}{|M{0.48\linewidth}|}{Projected subspace $\projS_+$} \\
\cline{2-6}
 & $e_1(\beta)$ & $e_1(\beta)$ & $e_2(\beta)$ & $e_3(\beta)$ & $e_4(\beta)$ \\
\MidRule
2.0 & -42.6405(5) & -42.95(2) & -42.54(2) & -42.25(14) & -40.1(4) \\
3.0 & -42.8159(2) & -43.11(6) & -42.75(6) & -42.60(8) & -40.8(8) \\
5.0 & -42.9406(2) & -43.11(1) & -42.852(3) & -42.63(8) & - \\
10.0 & -43.0372(2) & -43.0998(3) & -42.9157(5) & - & - \\
15.0 & -43.0724(3) & -43.1036(3) & -42.927(1) & - & - \\
\BotRule
\end{tabular}
\caption{Projected thermal energies obtained using our Monte Carlo method in the quantum Heisenberg limit for $L=8$ at various values of $\beta$. The second column gives the ground-state energy obtained using $\projS_{(1)}$. Because this subspace is free of sign problems, the single thermal energy $e_1(\beta)$ can be extracted accurately even at large $\beta$. The remaining four columns give the projected thermal energies obtained from $\projS_+$. Due to sign problems which sneak in from the larger projection subspace, we had to drop the projection state $|\psi_4\rangle$ at $\beta=5$, and used only $|\psi_1\rangle$ and $|\psi_2\rangle$ at $\beta=10,15$. These results can be compared with the ground state energy of $-43.103168(256)$  in \cite{PhysRevB.56.11678}, the first excited state energy of $-42.90(4)$ and the second excited state energy of $-42.54(3)$ both obtained in \cite{PhysRevB.49.5983}.
\label{tab:L8GS}}
\end{table*}

\begin{figure*}
\centering
\includegraphics[width=0.6\linewidth]{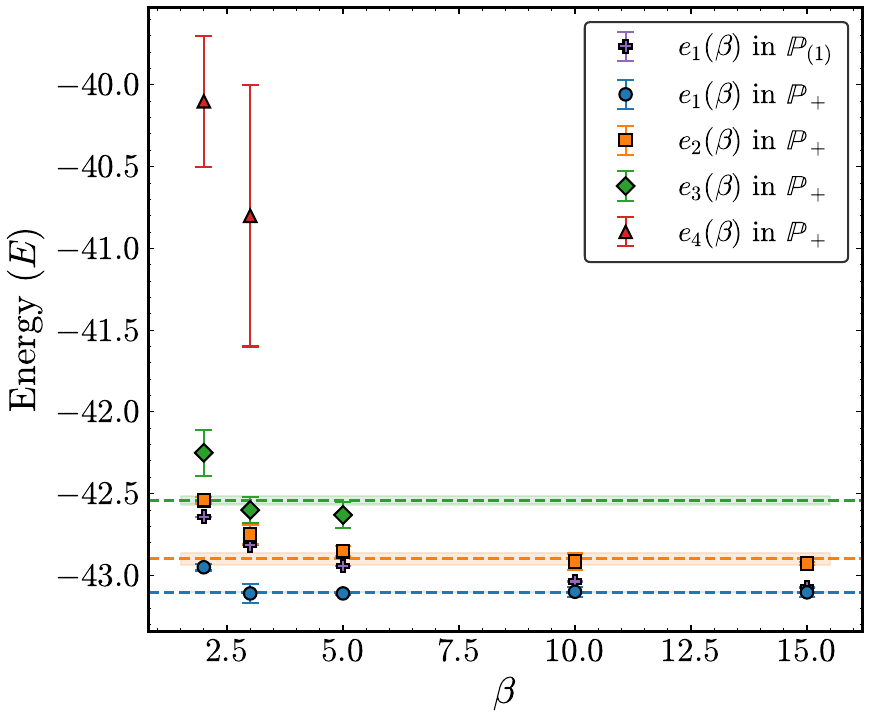}
\caption{
Projected thermal energies for the 2D Heisenberg square lattice, at \( L = 8 \) 
as functions of inverse temperature \( \beta \). The plus (purple) symbol shows the ground-state thermal energy \( e_{1}(\beta) \) obtained within the single-state projection subspace \( \mathbb{P}_{(1)} \), which is free of sign problems and thus yields accurate estimates even at large \( \beta \). However, the results from this does not converge to the ground state (the lowest horizontal dashed line obtained in \cite{PhysRevB.56.11678}). The other symbols display the projected thermal energies \( e_{i}(\beta) \) (\( i = 1, \dots, 4 \)) obtained from the full $\projS_+$. The circle symbol gives
\( e_1(\beta)\), which is already consistent with the ground state energy at $\beta=3.0$. To avoid computational instabilities the projection state \( |\psi_{4}\rangle \) was excluded from the analysis at \( \beta = 5 \), and only \( |\psi_{1}\rangle \) and \( |\psi_{2}\rangle \) were used at \( \beta = 10, 15 \). The higher two horizontal dashed lines are reference Monte Carlo results with their uncertainties (shaded bands) from the earlier study of Ref.~\cite{PhysRevB.49.5983}. These results demonstrate that (i) \ac{PDMS} results converge to the ground state at very low $\beta$; (ii) Comparing the blue circles with the purple plus markers, we learn that increasing the dimension of the projection subspace judiciously can considerably improve the rate of convergence.}
\label{fig:2D-Heisenberg-L=8}
\end{figure*}

It would be interesting to explore alternative projection subspaces guided by both spin and lattice symmetries. One manifestation of these symmetries appears in the allowed degeneracies $g$ of the spin representations. For example, from \cref{tab:L4evals}, we observe that for $J = 1$, the degeneracies take the values $g = 1, 2, 4$, whereas for $J = 0$, they are $g = 1, 3, 4, 6, 8$. A more detailed understanding of these degeneracies, and the use of this information to design improved projection subspaces, could lead to more efficient and accurate implementations of our method.

\begin{figure}
\centering
\includegraphics[width=0.9\columnwidth]{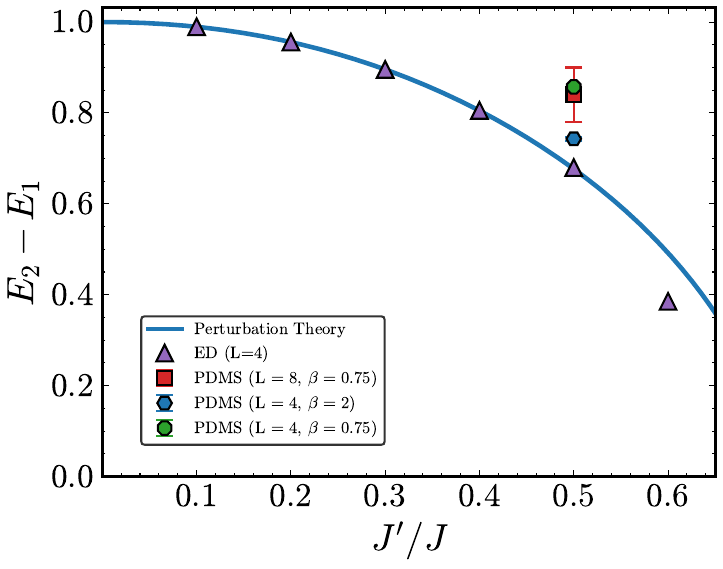}
\caption{ The energy gap between the ground state (with energy $E_1$) and the first excited Triplet state (with energy $E_2$) plotted as a function of the \ coupling $J'/J$. The solid blue is obtained using perturbation theory up to $(J'/J)^{21}$ from Ref. \cite{PhysRevB.65.014408}. The colored markers without errors are ED results from a $4\times4$ lattice. The agreement between ED and perturbative results confirms that non-perturbative effects remain small up to $J'/J\approx0.5$. The data from our Monte Carlo method at $J'/J=0.5$ for $L=4$ (that also appear in \cref{fig:L4plots}) are shown to demonstrate the rate of convergence. We also show data from our Monte Carlo method at $J'/J=0.5$, $L=8$ and $\beta=0.75$ and find that it agrees with the $L=4$ data within errors.
}
\label{fig:gap_vs_Jp}
\end{figure}

\subsection{The Quantum Heisenberg Limit}
\label{sec6-F}

We now investigate the performance of our \ac{PDMS} method on an $L = 8$ lattice, which lies beyond the reach of exact diagonalization studies. We first focus on the case $(J = 0, J' = 1)$, which turns the SSM into the well studied two dimensional quantum Heisenberg model. In this study we focus on the $\projS_+$ subspace. In the quantum Heisenberg limit, we can show that $Z_{11}/Z^b = 1$ for all values of $\beta$ and $L$, a result that can be easily proved, for example, using meron-cluster properties~\cite{Chandrasekharan:1999cm}. This implies that the sign problem is absent, at least for the projection state $|\psi_1\rangle$ at this coupling.

In this limit, our Monte Carlo method therefore enables a high-precision determination of $e_1(\beta)$ associated with the projection subspace $\projS_{(1)}$ spanned by $|\psi_1\rangle$. At large values of $\beta$, this projected thermal energy is expected to yield the ground-state energy of the two-dimensional Heisenberg antiferromagnet. Results for the $L = 8$ lattice are shown in the second column of \cref{tab:L8GS} and also plotted in \cref{fig:2D-Heisenberg-L=8}. We observe that $e_1(\beta)$ decreases gradually with increasing $\beta$. A previous study 
performed at $\beta = 64$ reported the ground-state energy to be $-43.103168(250)$~\cite{PhysRevB.56.11678}, whereas our current calculation yields, $e_1(\beta = 15) = -43.0724(3)$, which has clearly not converged to the low temperature limit.

Rather than taking the large-$\beta$ limit with $\projS_{(1)}$, we can instead use $\projS_+$ to test whether increasing the dimension of the projection subspace improves the rate of convergence. The projected thermal energies obtained using $\projS_+$ are listed in the right four columns of \cref{tab:L8GS} and also plotted in \cref{fig:2D-Heisenberg-L=8}.
We now find $e_1(\beta = 3.0) = -43.11(6)$ and $e_1(\beta = 5.0) = -43.11(1)$, both showing clear signs of convergence to the ground state result of \cite{PhysRevB.56.11678}. The fact that we can obtain the ground state energy with an error of less than $0.2$\% even at the value of $\beta=3.0$ is impressive.

We note that the Monte Carlo errors in the calculation of $e_1(\beta)$ arise from contributions across the entire projection subspace. Sign problems can reemerge through other projection states for which $Z_{aa}/Z^b$ becomes small as $\beta$ increases. The use of larger projection subspaces therefore typically restricts us to smaller values of $\beta$. We show this behavior in \cref{tab:L8GS}. At $\beta = 3.0$, we are able to extract all four $e_a(\beta)$ values, but at $\beta = 5.0$ we had to discard the state $|\psi_4\rangle$ to obtain reliable results due to zero modes creeping in from that state. However, this reduction still allows us to recover the three lowest levels with equal or higher accuracy. Notably, $\dproj = 3$ performs better than $\dproj = 1$ for the ground-state calculation at $\beta = 5.0$. For $\beta = 10$ and $15$, the projection subspace had to be reduced further. By retaining only $|\psi_1\rangle$ and $|\psi_2\rangle$, we obtained more accurate estimates for the ground-state energy, $-43.1036(3)$, and the first excited-state energy, $-42.927(1)$. This estimate of the ground-state energy is consistent with that obtained at $\beta = 5.0$ with $\dproj = 3$. Another earlier work (see \cite{PhysRevB.49.5983}) reported the first excited energy as $-42.90(4)$, consistent with our result but with substantially larger errors. For the second excited state, the previous estimate of $-42.54(3)$ and our result of $-42.63(8)$ at $\beta = 3$ are also consistent, although our uncertainties are larger. We expect that a more carefully chosen projection subspace could improve the precision of some the higher excited states further.

\subsection{Perturbative Energy Gap at small \texorpdfstring{$J'/J$}{J'/J}}
\label{sec6-G}

In \cref{sec6-B} we explained that $\projS_-$ is an optimal subspace for computing the first excited energy level for small values of $J'/J$. Here we verify this by exploring how the $\projS_-$ subspace performs at $(J = 1, J' = 0.5)$. The energy gap between the ground state ($E_1$) and the first excited state ($E_2$) was computed in perturbation theory to very high orders  \cite{PhysRevB.65.014408} and the first few terms of this series are given by
\begin{align}
(E_2-E_1)\ =\ 1 - (\frac{J'}{J})^2 -  
- \frac{1}{2}(\frac{J'}{J})^3 
- \frac{1}{8}(\frac{J'}{J})^4 + \ldots.
\end{align}
In \cref{fig:gap_vs_Jp}, we plot the perturbation theory result for the gap as a function of $J'/J$. For comparison we
also show the results for the gap obtained using exact diagonalization on a $L=4$ lattice. The Monte Carlo results presented in \cref{fig:L4plots} for $L=4$, $J'/J = 0.5$ at $\beta=0.75$ and $\beta=2$ are also shown. These results suggest that the energy gap saturates even at $L=4$ and going to larger lattices will give results similar to the $L=4$ results. We have verified this at $L=8$.

For $L = 8$, the ground state is the dimer state of \cref{eq:dimerstate}, with energy $E_{\mathrm{GS}} = -3J L^2 / 8$ which means $E_{\mathrm{GS}} = -24$. We now examine whether we can compute the energy gap to the first excited state using the $\projS_-$ subspace, which now has $\dproj = 32$  as explained in \cref{sec6-B}, should work well. Similar to the $L = 2$ and $L = 4$ cases, we again find empirically that $\langle \phi_a | e^{-\beta H} | \phi_c \rangle \propto \delta_{ac}$ to a good approximation in our Monte Carlo calculations, although now the sign problem manifests as large fluctuations even at $\beta = 1.0$. This indicates that the projected thermal energies obtained within this high-dimensional subspace are degenerate within Monte Carlo errors. We used a four-dimensional subspace within $\projS_-$ to compute this degenerate thermal energy levels, which should converge to the first excited-state energy in the limit of large $\beta$. We find it to be $-23.16(6)$ at $\beta = 0.75$. This results is also shown in \cref{fig:gap_vs_Jp}. We see that $L=4$ and $L=8$ data are identical within errors as expected.

\section{Conclusions}
\label{sec7}

In this work, we introduced a new Monte Carlo sampling approach for computing the projected low-energy spectrum of a general quantum Hamiltonian $H$. Our approach targets the subset of low-energy eigenstates $|\alpha\rangle$ of $H$ that have nonzero overlap with a chosen $\dproj$-dimensional projection subspace $\projS$, spanned by linearly independent states $|\psi_a\rangle$ ($a = 1, 2, \ldots, \dproj$). As discussed in \cref{sec4}, this formulation enables the systematic identification of the projected low-energy eigenvalues ${\cal E}_{\alpha_a}$ directly from $H$ and the chosen subspace. This provides a versatile numerical strategy that can be tailored to diverse quantum many-body systems.

Our Monte Carlo method enables the computation of the projected thermal energies $e_a(\beta)$, which smoothly approach the projected low-energy eigenvalues ${\cal E}_{\alpha_a}$ in the $\beta \rightarrow \infty$ limit. The method operates by evaluating two $\dproj \times \dproj$ matrices: the projected thermal density matrix $Z$ and the corresponding projected thermal energy matrix $E$. All matrix elements are estimated through our Monte Carlo sampling procedure up to a common normalization factor, as is typical in such approaches. The projected thermal energies $e_a(\beta)$ are then obtained as the eigenvalues of the matrix product $E Z^{-1}$. We refer to this approach as the \ac{PDMS} method.

Building on earlier concepts, our work establishes a systematic and broadly applicable framework that significantly extends previous Monte Carlo approaches. First, we sample the density matrix $e^{-\beta H}$ in continuous Euclidean time within the path-integral formulation, starting from a lattice quantum Hamiltonian in a finite Hilbert space. This eliminates Trotter errors and allows flexible exploration of different projection subspaces. Second, as demonstrated in this study, our method remains largely insensitive to sign problems at small values of $\beta$, enabling accurate evaluation of $e_a(\beta)$ even in highly frustrated regimes of the Shastry--Sutherland model (SSM). This robustness makes the technique suitable for a wide class of systems. Third, we show that with a judicious choice of projection subspaces, the projected thermal energies converge rapidly to the projected low-energy spectrum by providing reliable and accurate results even at small values of $\beta$.

 As discussed in \cref{sec4}, both the precision and the rate of convergence of $e_a(\beta)$ toward the projected low-energy spectrum of $H$ can be systematically improved, at fixed $\beta$, by carefully expanding the dimension $\dproj$ of the projection subspace. We also introduced the concept of an ideal projection subspace for extracting a chosen set of $\dproj$ projected low-energy levels. Such a subspace is characterized by the property that the corresponding projected eigenvectors of $H$ form $\dproj$ linearly independent vectors within the subspace, while all remaining projected eigenvectors have strictly higher energies. This idea provides a practical guideline for constructing efficient and well-targeted projection bases in future applications.

Using these analytic insights, we demonstrated in \cref{sec5,sec6} that our method remains robust across a wide range of projection subspaces at high temperatures. In practice, we find that stable convergence close to the projected low-energy spectrum can often be achieved at moderate values of $\beta$, without encountering severe sign problems. At larger $\beta$ or with significantly expanded projection subspaces, sign problems may reemerge, similar to the challenges faced when extracting higher-energy excitations in otherwise sign-problem-free systems~\cite{Wagman:2017xfh}. These observations highlight the importance of carefully selecting projection subspaces that target the desired low-energy levels efficiently. We demonstrated this strategy successfully in our study of the transverse-field Ising model and $J=0$ regime of the SSM, where well-chosen projection states enabled rapid and stable convergence.

A further strength of the \ac{PDMS} framework is that the Monte Carlo sampling can, in principle, be formulated in a variety of bases within the Hilbert space. In some cases, sign problems can be completely resolved through a careful choice of basis~\cite{PhysRevB.57.R3197,PhysRevLett.117.197203}, while in others they can be significantly mitigated~\cite{JensEisert2020}. It may also be possible to take guidance from analytic techniques such as perturbation theory when constructing the sampling basis. Furthermore, new ideas for addressing sign problems, such as meron clusters~\cite{Chandrasekharan:1999cm}, fermion bags~\cite{Chandrasekharan:2013rpa}, exponential error reduction techniques using the multi-level blocking ideas \cite{Luscher:2001up,Mak:1998zw,Yoo:2004mh}, can all be naturally incorporated into our framework. These features showcase the broad adaptability of the method and its potential for integration with parallel advances in Monte Carlo methodology.

An interesting avenue for further development is the use of physically-motivated construction of projection states that are guided by the known quantum phases of the underlying Hamiltonian, as demonstrated in recent tensor-network studies~\cite{Corboz2025}. For example, projection subspaces inspired by phase-specific wavefunctions, such as plaquette-valence-bond (PVB) type states in the Shastry--Sutherland model, may provide more efficient starting points for our \ac{PDMS} method. Our preliminary results show that such physically motivated choices may considerably accelerate convergence and improve stability in challenging regimes. We plan to share these findings in a future work.

The \ac{PDMS} framework naturally admits a physical interpretation in terms of an effective thermal system defined by the projected thermal density matrix. The corresponding effective energies $e_a(\beta)$ characterize this system at finite temperature, while their $\beta \rightarrow \infty$ limit yields the projected low-energy spectrum of $H$. Viewed in this way, our method provides a nonperturbative and efficient approach to exploring this effective thermal problem at high temperatures, where Monte Carlo simulations are most stable. Moreover, developing an analytic understanding of the $\beta$ dependence of $e_a(\beta)$ may enable systematic extrapolations from high-temperature results to the low-temperature regime. This perspective is conceptually similar to finite-size scaling analyses, where controlled extrapolations from smaller systems can predict behavior in the thermodynamic limit. Establishing this connection more concretely presents a natural direction for future research.

Thus, by combining the aforementioned features and directions for further refinement,  we believe that the \ac{PDMS} method emerges as a powerful and versatile computational framework for determining the low-energy spectrum of generic quantum Hamiltonians. Its flexibility in defining projection subspaces, robustness against sign problems at low $\beta$, and compatibility with modern Monte Carlo and analytical techniques make it broadly relevant across diverse problems in quantum many-body physics. We anticipate that the method will find applications not only in condensed-matter systems, such as frustrated magnets and quantum spin models demonstrated here, but also in lattice gauge theories formulated in the Hamiltonian framework. More broadly, we hope that this work will encourage the development of projection-based Monte Carlo strategies as a general paradigm for exploring complex quantum systems beyond the reach of traditional approaches.

\section*{Acknowledgments}

SC thanks Subhro Bhattacharjee, Souvik Kundu, and Indrajit Sau for collaboration on another project involving the \ac{PDMS} method; discussions from that work have also influenced the progress reported here. SC is supported in part by the U.S. Department of Energy, Office of Science, Nuclear Physics program under Award No.~DE-FG02-05ER41368. RKK and HSW were supported by the NSF under Award No. DMR-2312742. SC was also supported in part by the International Centre for Theoretical Sciences (ICTS) under its Associates Program, during which some of the ideas of the present work were developed. AK thanks Sara Haravifard and Lalit Yadav for introducing the exciting physics of the Shastry-Sutherland Model. AK is supported by funds from Duke University for this research.

\appendix

\begin{table*}[t]
\centering
\renewcommand{\arraystretch}{1.15}
\begin{tabular}{|r||r|l||r|l||r|l||r|l|}
\hline
\multicolumn{1}{|c||}{$r$} & \multicolumn{1}{c|}{$\langle \psi|\alpha_0\rangle$} & \multicolumn{1}{c||}{$\psi$-rep} &
\multicolumn{1}{c|}{$\langle \psi|\alpha_1\rangle$} & \multicolumn{1}{c||}{$\psi$-rep} &
\multicolumn{1}{c|}{$\langle \psi|\alpha_2\rangle$} & \multicolumn{1}{c||}{$\psi$-rep} &
\multicolumn{1}{c|}{$\langle \psi|\alpha_3\rangle$} & \multicolumn{1}{c|}{$\psi$-rep} \\
\hline
1 & 0.6908 & $\uparrow\,\uparrow\,\uparrow\,\uparrow\,\uparrow\,\uparrow\,\uparrow\,\uparrow\,\uparrow\,\uparrow\,\uparrow\,\uparrow$ &
  0.3576 & $\uparrow\,\uparrow\,\uparrow\,\uparrow\,\uparrow\,\uparrow\,\uparrow\,\uparrow\,\uparrow\,\uparrow\,\uparrow\,\uparrow$ &
  0.3031 & $\downarrow\,\downarrow\,\downarrow\,\downarrow\,\downarrow\,\downarrow\,\downarrow\,\downarrow\,\downarrow\,\downarrow\,\downarrow\,\uparrow$ &
  0.4027 & $\downarrow\,\downarrow\,\downarrow\,\downarrow\,\downarrow\,\downarrow\,\downarrow\,\downarrow\,\downarrow\,\downarrow\,\downarrow\,\downarrow$ \\
\hline
2 & 0.5632 & $\downarrow\,\uparrow\,\uparrow\,\uparrow\,\uparrow\,\uparrow\,\uparrow\,\uparrow\,\uparrow\,\uparrow\,\uparrow\,\uparrow$ &
  0.3356 & $\downarrow\,\downarrow\,\downarrow\,\uparrow\,\uparrow\,\uparrow\,\uparrow\,\uparrow\,\uparrow\,\uparrow\,\uparrow\,\uparrow$ &
  0.2604 & $\uparrow\,\uparrow\,\uparrow\,\uparrow\,\uparrow\,\uparrow\,\uparrow\,\uparrow\,\uparrow\,\uparrow\,\uparrow\,\uparrow$ &
  0.3979 & $\downarrow\,\downarrow\,\downarrow\,\downarrow\,\downarrow\,\downarrow\,\downarrow\,\downarrow\,\downarrow\,\downarrow\,\downarrow\,\uparrow$ \\
\hline
3 & 0.2371 & $\downarrow\,\downarrow\,\uparrow\,\uparrow\,\uparrow\,\uparrow\,\uparrow\,\uparrow\,\uparrow\,\uparrow\,\uparrow\,\uparrow$ &
  0.3249 & $\downarrow\,\downarrow\,\downarrow\,\downarrow\,\uparrow\,\uparrow\,\uparrow\,\uparrow\,\uparrow\,\uparrow\,\uparrow\,\uparrow$ &
  0.2576 & $\downarrow\,\downarrow\,\downarrow\,\downarrow\,\downarrow\,\downarrow\,\downarrow\,\downarrow\,\downarrow\,\downarrow\,\downarrow\,\downarrow$ &
  0.3004 & $\downarrow\,\downarrow\,\downarrow\,\downarrow\,\downarrow\,\uparrow\,\uparrow\,\uparrow\,\uparrow\,\uparrow\,\uparrow\,\uparrow$ \\
\hline
4 & 0.1363 & $\downarrow\,\uparrow\,\downarrow\,\uparrow\,\uparrow\,\uparrow\,\uparrow\,\uparrow\,\uparrow\,\uparrow\,\uparrow\,\uparrow$ &
  0.2701 & $\downarrow\,\downarrow\,\downarrow\,\downarrow\,\downarrow\,\uparrow\,\uparrow\,\uparrow\,\uparrow\,\uparrow\,\uparrow\,\uparrow$ &
  0.2539 & $\downarrow\,\downarrow\,\uparrow\,\uparrow\,\uparrow\,\uparrow\,\uparrow\,\uparrow\,\uparrow\,\uparrow\,\uparrow\,\uparrow$ &
  0.2884 & $\downarrow\,\downarrow\,\downarrow\,\downarrow\,\downarrow\,\downarrow\,\uparrow\,\uparrow\,\uparrow\,\uparrow\,\uparrow\,\uparrow$ \\
\hline
\end{tabular}
\caption{The zero-momentum $Z$-basis states $|\psi\rangle$ with the largest overlap with the energy eigenstates $|\alpha\rangle$. A representative for the $k=0$ state $|\psi\rangle$ is given, from which the full state is constructed via symmetrization. Only the $r=4$ state for $|\alpha_0\rangle$ lies outside the 0 and 2 domain wall sector.}
\label{tab:overlaps-top4}
\end{table*}

\section{Stochastic Series Expansion}
\label{app:sse}

While we chose to present our \ac{PDMS} algorithm using a path integral formulation, it can equally well be formulated using the popular stochastic series expansion (SSE)~\cite{Sandvik2010_QMCReview}. The steps are self evident but are included here for completeness. To compute the $Z$ matrix, we simply expand the exponential as usual in the SSE,
\begin{align}
Z_{ac} &= \langle \psi_a | e^{-\beta H} | \psi_c \rangle\\
&=\sum_n \langle \psi_a | \frac{ (-\beta H)^n}{n!} | \psi_c \rangle.
\label{eq:Zmat_SSE}
\end{align} 
and insert complete sets of states between each $H$ and then sample the various bond terms which $H$ can be written as a sum of, Eq.~\ref{eq:mcH}. The updates are standard, there are both diagonal and off-diagonal updates, though now one has to allow for the possibility of the boundary states $|\psi_a\rangle$ and $|\psi_c\rangle$ to be updated, along the lines of what we have described in the path integral approach. The matrix $E$ can also be constructed in this manner. Once both these matrices are constructed they can be analyzed exactly as described in our paper. Since the Monte Carlo sampling procedures are very similiar, it is anticipated that the SSE formulation will not offer any signifcant advantages or disadvantages as compared to the path integral formulation we have formulated here, indeed apart from some book-keeping, both methods are quite similar.

\section{Details: Transverse Field Ising Model}
In this appendix, we provide further details of our calculations related to the application of the \ac{PDMS} to the \ac{TFIM}.

\subsection{Projection subspace}

Our choice of the projection subspace, $\projS$, was made using data from exact diagonalization results on small lattices. We explored which Z-basis states have the highest overlap with the lowest energy eigenstates of $H$. In \cref{tab:overlaps-top4} we show the four largest overlaps between zero-momentum $Z$-basis states and the exact energy eigenstates for $L=12$ and $h=0.2$. We see that all but one state ($r=4$ for $|\alpha_{0}\rangle$) falls within the zero or two domain wall sectors. This motivated us to focus on states of the form $|\psi_{\ell+2}\rangle$ defined in \cref{eq:projTFIMl}. While there are clearly large overlaps with these states for $L/2 <\ell< L $, we only focused on the states where $1 <\ell \leq L/2$ in order to reduce the dimension of $\projS$. This restricted space continues to be a good choice. Motivated by the largest overlap seen for the fourth excited state we added the state \cref{eq:projTFIM2} where all the spins were flipped. On large lattices this state is clearly not necessary.

\subsection{Zero modes of \texorpdfstring{$Z$}{}}

The extraction of the projected thermal energies $e_{a}(\beta)$ relies on inverting the matrix $\hat{Z}$. While the exact eigenvalues of $\hat{Z}$ are guaranteed to be positive, this is not true in the presence of Monte Carlo error. Since the eigenvalue ratios are expected to be of the form  
$e^{-\beta (\mathcal{E}_{a}-\mathcal{E}_{0})}$, for higher energy modes at low temperatures, eigenvalues of $\hat{Z}$ can become indistinguishable from zero (within errors). In this case we encounter a ``zero mode'', obstructing the extraction of \textit{any} projected energies.

One na\"ive strategy is to take a smaller projection subspace with dimension approximately equal to the number of low energy modes we are targeting. However, we find that taking a small projection subspace drastically slows the convergence to the $\beta\rightarrow\infty$ energies, making this strategy unfeasible for extracting the projected low energy spectrum. Moreover, this does not necessarily mitigate the zero mode issue.

We have discovered that a better approach is to keep a large projection subspace, but then \textit{further project} into the subspace of $\hat{Z}$ that is zero-mode free as we explained in \cref{sec4}. In this approach we diagonalize $Z/Z^b$ and then project both $Z/Z^b$ and $\tilde{E}/Z^b$ to a lower dimensional subspace of dimension $\dprojres$ such that $Z/Z^b$ is strictly positive in this lower dimensional subspace. The projected thermal energies $e_{a}(\beta)$ are then well defined.

\begin{figure}
    \centering
    \includegraphics[width=1.0\linewidth]{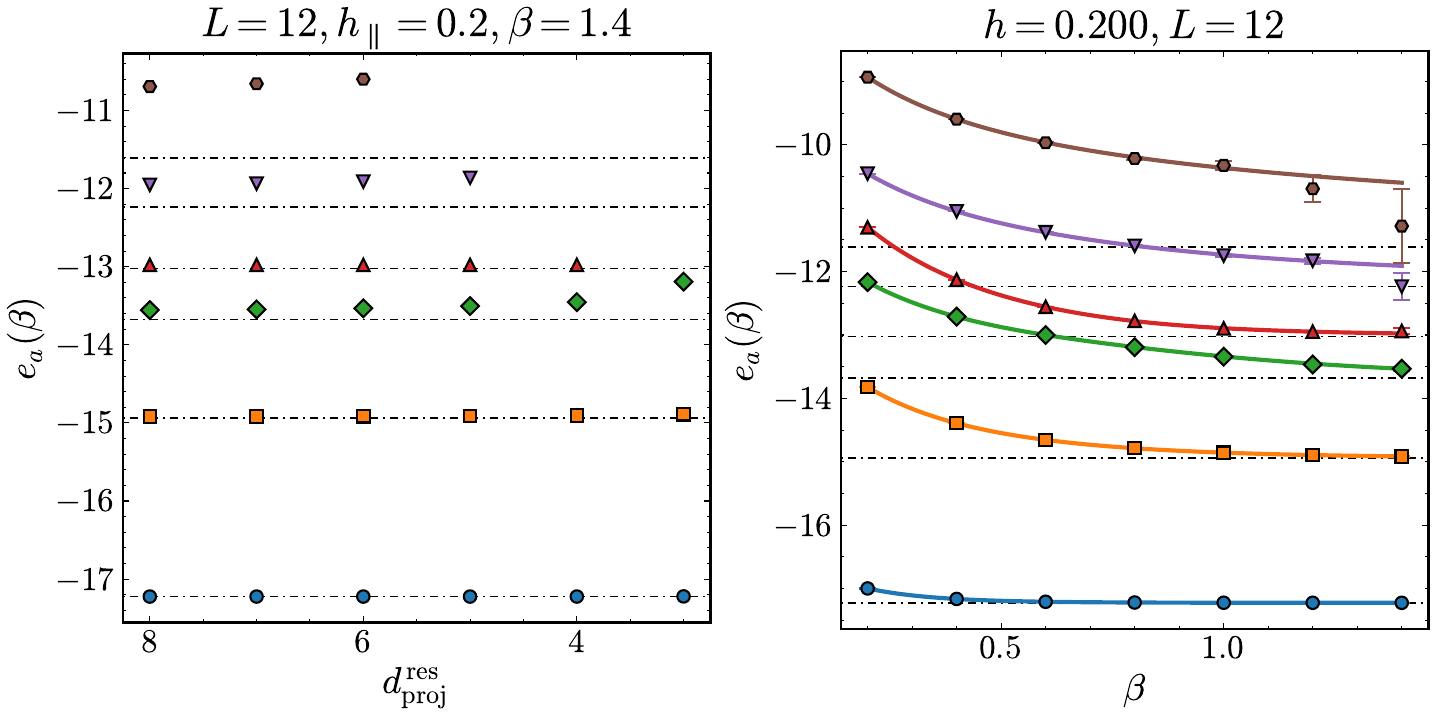}
    \caption{(a) Exact diagonalization study of projected energies as a function of $d_{\mathrm{proj}}^{\mathrm{res}}$ for a fixed projected subspace of dimension $d_{\mathrm{proj}}=8$ (b) \ac{PDMS} energies as a function of $\beta$ with $d^{\mathrm{res}}_{\mathrm{proj}}=6$ and $d_{\mathrm{proj}}=8$.}  
    \label{fig:zero_mode_proj}
\end{figure}

A natural question to ask is how the projection from the original $\dproj$ dimensional subspace to the smaller $\dprojres$ subspace effects the projected thermal energies. We have explored this question by comparing exact diagonalization results with results obtained from Monte Carlo at $L=12$ and with a $\dproj=8$. In the left panel of \cref{fig:zero_mode_proj}, we  show the exact diagonalization results for the effect of lowering $\dprojres$. We find that the effect is relatively mild with the projected energies staying close to the zero temperature results (dashed lines). In the right panel of \cref{fig:zero_mode_proj} we plot the results for the projected thermal energies as a function of $\beta$ obtained using the Monte Carlo method,
and focus on $\dprojres=6$. We observe that the Monte Carlo errors increase with $\beta$ for the highest two energies and this makes it difficult to extract reliable results at $\beta=1.4$.
If pushed to higher $\beta$, the $Z/Z^b$ will encounter a zero mode, obstructing energy extraction. In contrast we can even reach $\beta = 1.8$ if we use  $\dprojres=4$. This shows that if $\dprojres$ is too high, small eigenvalues arising from higher energy states can substantially contribute to the errors in the lower energy levels.

\label{app1}
\begin{figure*}[thb]
\centering
\includegraphics[width=0.8\columnwidth]{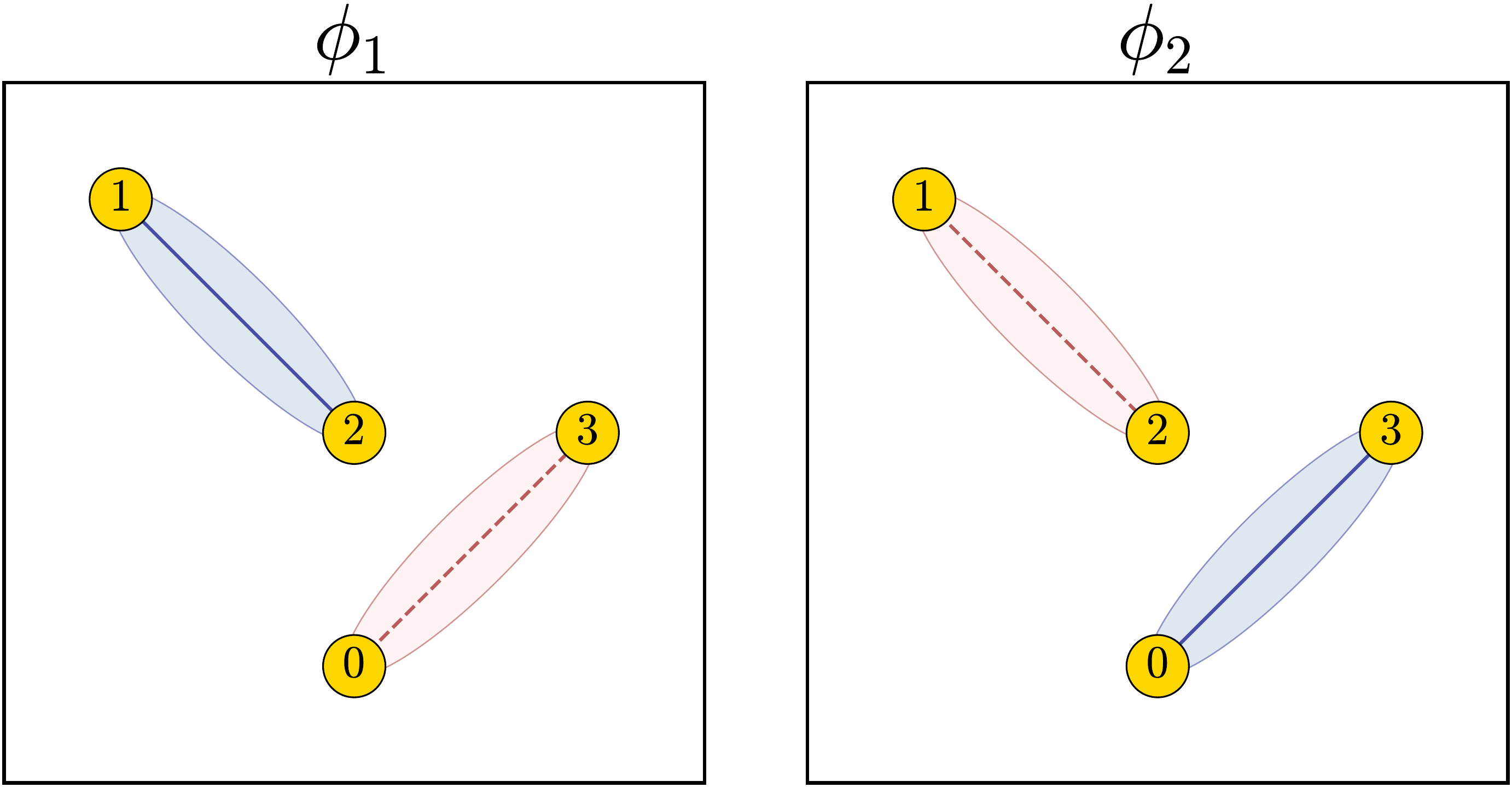}
\hspace{0.6in}
\includegraphics[width=0.8\textwidth]{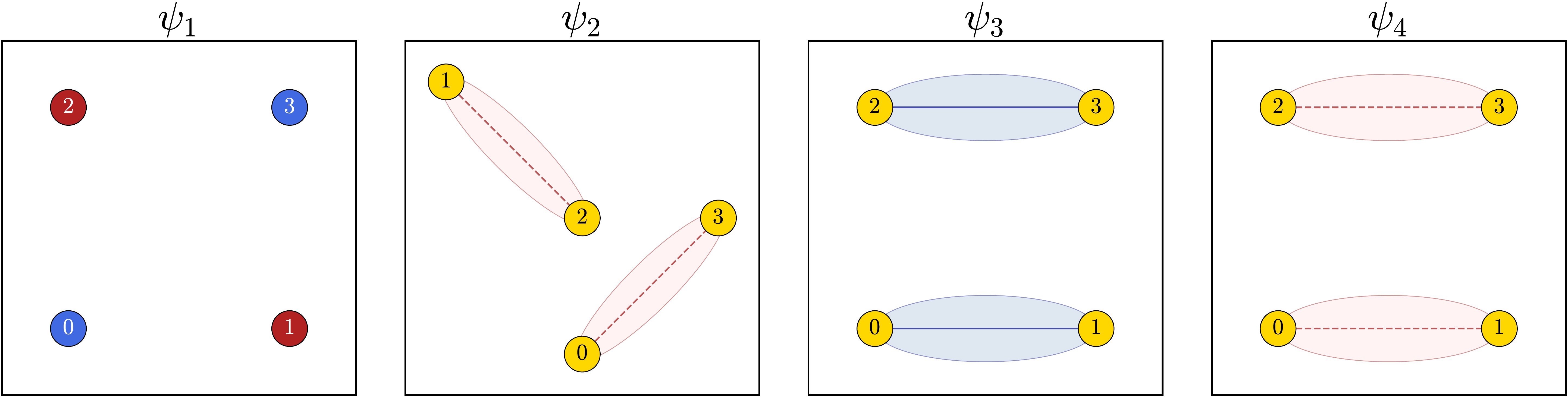}
\caption{The top two panels show the two projection states in the $\projS_-$ subspace on a $L=2$ lattice. The bottom four panels show the four projection states in the $\projS_+$ subspace on a $L=2$ lattice. The blue circle denote the state $(|\ua\rangle+|\da\rangle)$, the red circle denotes the state $(|\da\rangle-|\ua\rangle)$, the dashed dimers represent $|\psi_t\rangle$, and the solid dimers represent $|\psi_s\rangle$ as defined in \cref{eq:singlet,eq:triplet}.
}
\label{fig:2x2projst}
\end{figure*}

\begin{table*}[t]
\centering
\renewcommand{\arraystretch}{1.4}
\setlength{\tabcolsep}{4pt}
\begin{tabular}{|c|c|c|c|c|}
\TopRule
\multirow{2}{*}{index ($a$)}
 &
\multicolumn{4}{c|}{$E_a(s)(g)$} \\ \cline{2-5}
& ($J=0$, $J'=1$) & ($J=1$, $J'=0.5$) & ($J=1$, $J'=0.66$) & ($J=1$, $J'=0.75$) \\
\MidRule
\hline
1 & -4.00(0)(1) & -1.50(0)(2) & -2.14(0)(1) & -2.50(0)(1) \\
2 & -2.00(1)(1) & -0.50(1)(3) & -1.50(0)(1) & -1.50(0)(1) \\
3 & 0.00(0)(1) & 1.50(2)(1) & -0.82(1)(1) & -1.00(1)(1) \\
4 & 0.00(1)(2) &  & -0.50(1)(2) & -0.50(1)(2) \\
5 & 2.00(2)(1) &  & 1.82(2)(1) & 2.00(2)(1) \\
\hline
\BotRule 
\end{tabular}
\caption{Distinct energy levels of $H$ for $L=2$, listed in ascending order at four different coupling values. The leftmost column gives the index $a$ for each distinct energy level $E_a$. The remaining four columns show $E_a(s)(g)$ at the corresponding couplings $(J,J')$. Here $s$ denotes the spin representation and $g$ is the degeneracy associated with that representation. The total degeneracy of level $E_a$ is therefore $g(2s+1)$.}
\label{tab:L2evals}
\end{table*}

\section{Details: Shastry Sutherland Model}
\label{app2}

In this appendix, we provide details related to \cref{sec5} to help a reader follow the discussion.

\subsection{Lattice Details}

We first define the $\nn$ and $\nnn$ bonds that appear in the Hamiltonian more carefully. This helps us set up the notation that we use when to define the projection subspaces. We label each lattice site by its coordinates $(x_i, y_i)$, where $x_i, y_i = 0, 1, \ldots, L-1$. Sites for which $x_i + y_i$ is even (odd) are referred to as even (odd) sites. The $\nn$ bonds denote nearest-neighbor pairs, where we always choose $i$ to be an even site and $j$ to be the corresponding odd neighbor. The $\nnn$ bonds form a specific set of diagonal pairs, where both $i$ and $j$ are either even sites ($\nnn_A$ bonds) or odd sites ($\nnn_B$ bonds). In the $\nnn_A$ case, $i$ has even $x_i$, and $j$ is chosen with coordinates $x_j = x_i + 1$ and $y_j = y_i + 1$. In the $\nnn_B$ case, $i$ again has even $x_i$, while $j$ has $x_j = x_i - 1$ and $y_j = y_i + 1$. Periodic boundary conditions are imposed when coordinates cross the lattice boundaries. These definitions ensure that the $\nnn$ bonds partition the $L^2$ sites uniquely into two sets of $L^2 / 2$ bonds. All these features are illustrated on an $L=4$ lattice in \cref{fig:SSMlattice}.

\subsection{Definitions of local states}

We construct our projection subspaces using unnormalized singlet and triplet dimer states on the $\nn$ and $\nnn$ bonds. The singlet dimers are defined as
\begin{align}
|\psi_s\rangle &=
(-1)^{x_i+y_i}
\left(|\ua\rangle_i \otimes |\da\rangle_j
- |\da\rangle_i \otimes |\ua\rangle_j\right),
\label{eq:singlet}
\end{align}
and the triplet dimers are defined as
\begin{align}
|\psi_t\rangle &=
\left(|\ua\rangle_i \otimes |\da\rangle_j
+ |\da\rangle_i \otimes |\ua\rangle_j\right),
\label{eq:triplet}
\end{align}
where we assume $|\ua\rangle_i$ and $|\da\rangle_i$ to be eigenstates of the $s_i^z$ with eigenvalues $+1/2$ and $-1/2$. In the singlet definition, the phase factor $(-1)^{x_i+y_i}$ depends on site $i$, which was defined uniquely above when specifying the $\nn$ and $\nnn$ bonds.

\subsection{Exact results for \texorpdfstring{$L = 2$}{L = 2}}
\begin{figure*}[t]
\centering
\vbox{
\includegraphics[width=0.9\textwidth]{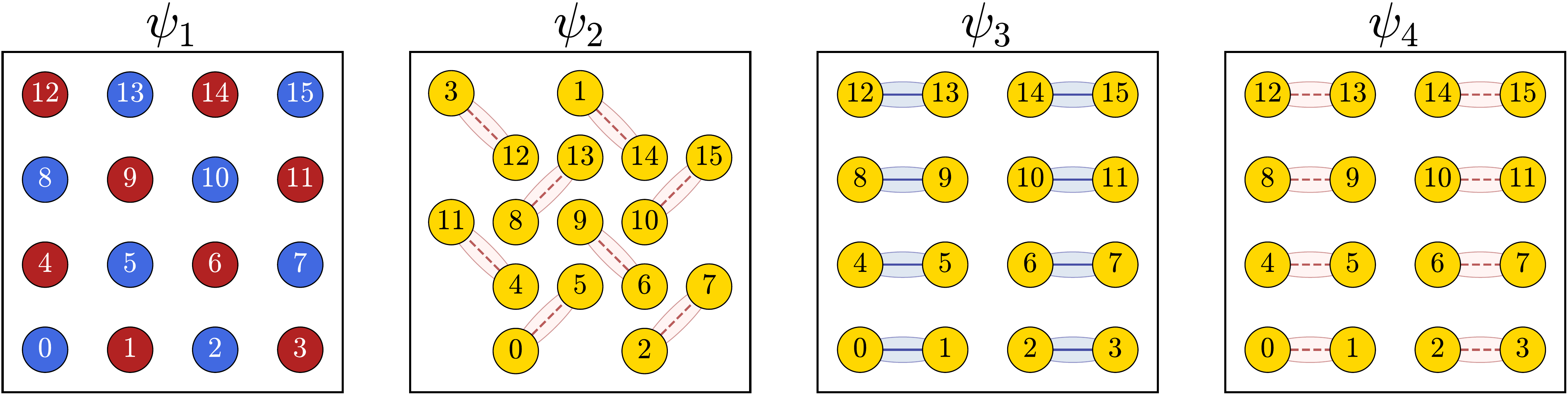} \\
\includegraphics[width=0.9\textwidth]{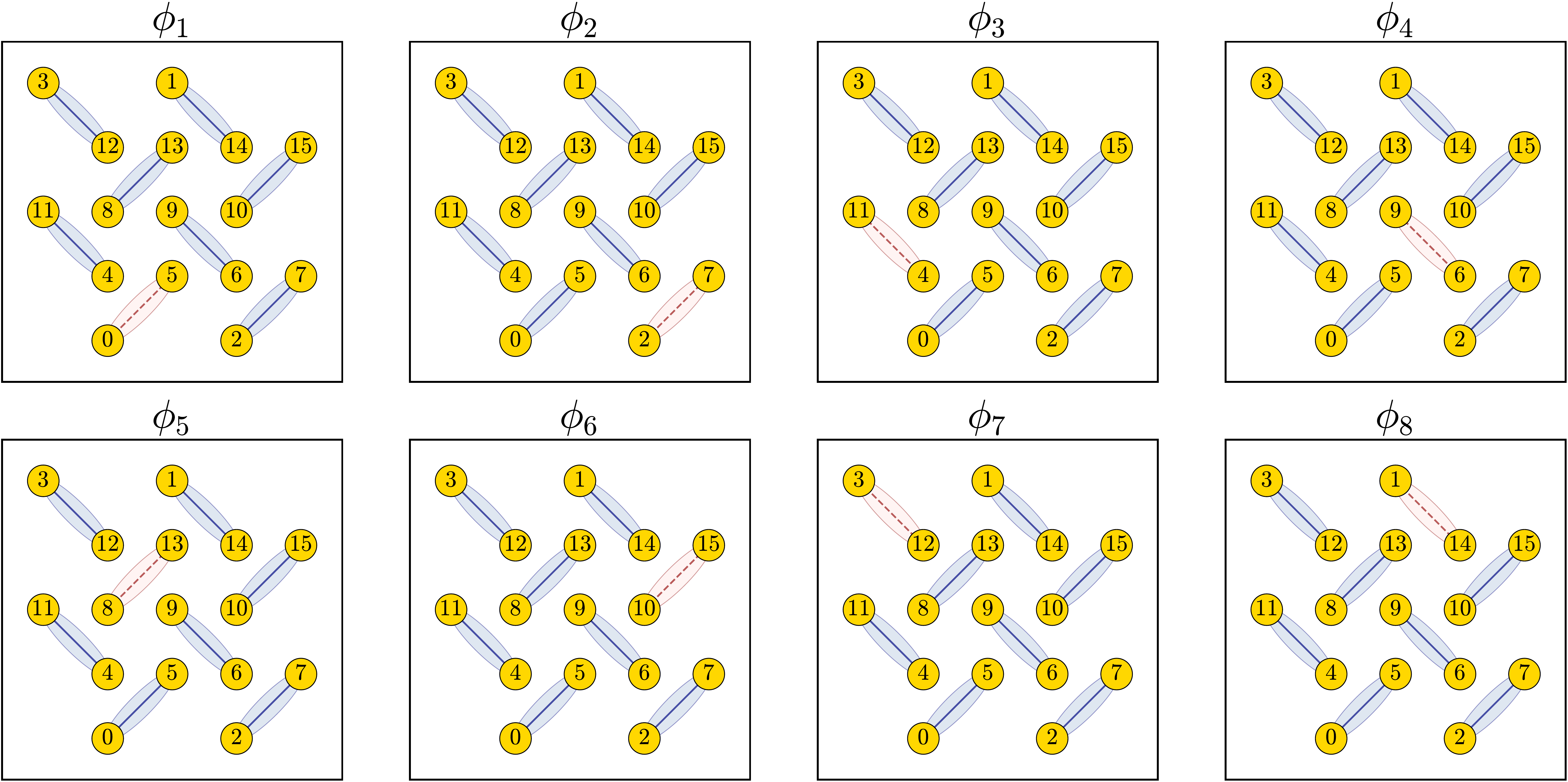}
}
\caption{The top row shows the four projection states in the 
$\projS_+$ subspace on a $L=4$ lattice. The bottom two rows show the 
the eight projection states in the $\projS_-$ subspace on a $L=4$ lattice. The definitions of the bonds are the same as in \cref{fig:2x2projst}.}
\label{fig:4x4projst}
\end{figure*}
For $L = 2$ we can use the dimer basis introduced in \cref{sec6} to construct the states. This dimer basis consists of a triplet state $|m\rangle$ ($m = \pm 1, 0$) or a singlet state $|s\rangle$ on each of the two diagonal bonds of the lattice. The first bond connects the sites with coordinates $(x, y) = (0, 0)$ and $(1, 1)$, while the second connects $(0, 1)$ and $(1, 0)$. Thus, the full $16$-dimensional Hilbert space in this basis consists of $9$ states labeled $|m; m'\rangle$, $3$ states labeled $|m; s\rangle$, $3$ labeled $|s; m\rangle$, and a single state $|s; s\rangle$.

In this dimer basis, the five $s = 2$ eigenstates of the SSM are
\begin{align}
|\chi_1\rangle &= |+1; +1\rangle, \\
|\chi_2\rangle &= \frac{1}{\sqrt{2}}\big(|+1; 0\rangle + |0; +1\rangle\big), \\
|\chi_3\rangle &= \frac{1}{\sqrt{6}}\big(|+1; -1\rangle + |-1; +1\rangle + 2|0; 0\rangle\big), \\
|\chi_4\rangle &= \frac{1}{\sqrt{2}}\big(|-1; 0\rangle + |0; -1\rangle\big), \\
|\chi_5\rangle &= |-1; -1\rangle,
\end{align}
and have energy $2J' + J/2$.

In addition, there are three $s = 1$ (triplet) eigenstates formed with triplets on both diagonal bonds:
\begin{align}
|\chi_6\rangle &= \frac{1}{\sqrt{2}}\big(|0; +1\rangle - |+1; 0\rangle\big), \\
|\chi_7\rangle &= \frac{1}{\sqrt{2}}\big(|+1; -1\rangle - |-1; +1\rangle\big), \\
|\chi_8\rangle &= \frac{1}{\sqrt{2}}\big(|0; -1\rangle - |-1; 0\rangle\big),
\end{align}
with energy $-2J' + J/2$.

The next two triplets are formed using one diagonal triplet and one diagonal singlet:
\begin{align}
|\chi_{10}\rangle &= |+1; s\rangle, &
|\chi_{11}\rangle &= |0; s\rangle, &
|\chi_{12}\rangle &= |-1; s\rangle, \\
|\chi_{13}\rangle &= |s; +1\rangle, &
|\chi_{14}\rangle &= |s; 0\rangle, &
|\chi_{15}\rangle &= |s; -1\rangle,
\end{align}
and both sets are degenerate with energy $-J/2$.

Finally, there are two singlets. The first is formed from two diagonal triplets:
\begin{align}
|\chi_9\rangle &= \frac{1}{\sqrt{3}}\big(|0; 0\rangle - |+1; -1\rangle - |-1; +1\rangle\big),
\end{align}
with energy $-4J' + J/2$, while the second is formed from two diagonal singlets:
\begin{align}
|\chi_{16}\rangle &= |s; s\rangle,
\end{align}
with energy $-3J/2$. These energies at the four couplings we study in this work are given in \cref{tab:L2evals}.

We can now compute the overlap of these energy eigenstates with the projection states spanning the $Q = +1$ and $Q = -1$ projection subspaces. The basis state $|\psi_1\rangle$ in the $Q = +1$ subspace is
\begin{align}
|\psi_1\rangle &= |+1; +1\rangle + |-1; -1\rangle - 2|0; 0\rangle + |-1; +1\rangle \nonumber \\
&\quad + \sqrt{2}\big(|0; +1\rangle - |+1; 0\rangle + |0; -1\rangle - |-1; 0\rangle\big),
\end{align}
which, when expressed in terms of the eigenstates of $H$, becomes
\begin{align}
|\psi_1\rangle &= |\chi_1\rangle + |\chi_5\rangle + 2(|\chi_6\rangle + |\chi_8\rangle) \nonumber \\
&\quad - \sqrt{\tfrac{2}{3}}\,|\chi_3\rangle - \tfrac{4}{\sqrt{3}}\,|\chi_9\rangle.
\end{align}

The other three states in the same subspace are
\begin{align}
|\psi_2\rangle &= 2|0; 0\rangle = \sqrt{\tfrac{8}{3}}\,|\chi_3\rangle + \tfrac{2}{\sqrt{3}}\,|\chi_9\rangle, \\
|\psi_3\rangle &= |+1; -1\rangle + |-1; +1\rangle - |0; 0\rangle + |s; s\rangle \nonumber \\
&= -\sqrt{3}\,|\chi_9\rangle + |\chi_{16}\rangle, \\
|\psi_4\rangle &= |+1; -1\rangle + |-1; +1\rangle + |0; 0\rangle - |s; s\rangle \nonumber \\
&= \tfrac{4}{\sqrt{6}}\,|\chi_3\rangle - \tfrac{1}{\sqrt{3}}\,|\chi_9\rangle - |\chi_{16}\rangle.
\end{align}

The two states in the $Q = -1$ projection subspace are
\begin{align}
|\phi_1\rangle &= |\psi_5\rangle = 2|0; s\rangle = 2|\chi_{11}\rangle, \\
|\phi_2\rangle &= 2|s; 0\rangle = 2|\chi_{14}\rangle.
\end{align}
Note that these are already exact eigenstates of $H$. We can use the above information to obtain the information given in \cref{tab:L2evals,tab:L2projE}. \cref{fig:2x2projst} gives a pictorial illustration of $\projS_+$ and $\projS_-$ subspaces for $L=2$.

\subsection{Exact Results for \texorpdfstring{$L = 4$}{L = 4}}

Here we list the exact energies obtained using exact diagonalization and 
the provide information about the overlap of the projection states $|\psi_1\rangle$, $|\psi_2\rangle$, $|\psi_3\rangle$ and $|\psi_4\rangle$, in the $\projS_+$ subspace and  $|\phi_1\rangle$ in the $\projS_-$ subspace with the energy eigenstates of $H$. The remaining seven projection states in the $\projS_-$ subspace have very similar projections as $|\phi_1\rangle$. 

Focusing first on the distinct energy levels of $H$, we use exact diagonalization to find them all. Although the total Hilbert space is considerably larger than the $L=2$ case, with dimension $2^{16} = 65{,}536$, we can restrict our calculation to the subspace of eigenstates of $H$ with total $s^z = 0$, which contains only $12{,}870$ states. This reduced Hamiltonian matrix can be diagonalized easily. Likewise, we can analyze each $s^z \neq 0$ sector separately, obtaining smaller matrices that can also be diagonalized, thereby yielding all eigenvalues and eigenvectors of $H$. We can compute $s$ by diagonalizing the $S^2$ operator in each degenerate subspace, since $[S^2,H]=0$. The first $45$ distinct levels that we find are listed in \cref{tab:L4evals}. \cref{fig:4x4projst} gives a pictorial illustration of $\projS_+$ and $\projS_-$ subspaces for $L=4$.

In order to find the projected low energy spectrum in the two subspaces, we next compute the overlap of the projection states with various energy eigenstates. In \cref{tab:L4ov-phi} we give the overlap of $|\phi_1\rangle$ and in \cref{tab:L4ov-psi} we give the overlap with $|\psi_1\rangle$, $|\psi_2\rangle$, $|\psi_3\rangle$ and $|\psi_4\rangle$.

\begin{table*}[t]
\centering
\renewcommand{\arraystretch}{1.4}
\setlength{\tabcolsep}{4pt}
\footnotesize
\begin{tabular}{|c|c|c|c|c|}
\TopRule
\multirow{2}{*}{index ($a$)}
 &
\multicolumn{4}{c|}{$E_a(s)(g)$} \\ \cline{2-5}
& ($J=0$, $J'=1$) & ($J=1$, $J'=0.5$) & ($J=1$, $J'=0.66$) & ($J=1$, $J'=0.75$) \\
\MidRule
\hline
1 & -11.228483(0)(1) & -6.000000(0)(1) & -6.000000(0)(1) & -6.884403(0)(1) \\
2 & -10.649885(1)(1) & -5.320287(1)(4) & -5.933096(0)(1) & -6.443055(0)(2) \\
3 & -9.517688(2)(1) & -5.320042(1)(4) & -5.916800(0)(2) & -6.370229(1)(1) \\
4 & -8.886442(1)(4) & -5.182600(0)(2) & -5.782810(0)(4) & -6.252202(0)(4) \\
5 & -8.794379(1)(4) & -5.117269(0)(4) & -5.645621(1)(4) & -6.152386(1)(4) \\
6 & -8.518284(1)(6) & -5.029625(0)(1) & -5.642281(0)(1) & -6.056300(1)(4) \\
7 & -8.252827(0)(3) & -5.027025(0)(1) & -5.639354(1)(4) & -6.043823(0)(1) \\
8 & -7.877905(0)(8) & -5.026103(0)(2) & -5.632751(0)(2) & -6.031847(0)(1) \\
9 & -7.836538(3)(1) & -5.019828(0)(2) & -5.628946(0)(1) & -6.022370(1)(4) \\
10 & -7.812148(0)(6) & -4.812839(0)(1) & -5.587025(0)(2) & -6.000000(0)(1) \\
11 & -7.765674(2)(4) & -4.804425(1)(4) & -5.506831(1)(4) & -5.999654(0)(2) \\
12 & -7.687234(0)(1) & -4.802876(1)(4) & -5.487639(1)(2) & -5.968997(0)(4) \\
13 & -7.630213(2)(4) & -4.793562(1)(2) & -5.439011(0)(1) & -5.964137(1)(2) \\
14 & -7.477053(2)(6) & -4.778225(1)(2) & -5.438800(1)(1) & -5.953098(0)(2) \\
15 & -7.432605(1)(6) & -4.769293(0)(4) & -5.422847(1)(4) & -5.895058(1)(4) \\
16 & -7.389958(1)(8) & -4.769273(0)(4) & -5.394965(1)(4) & -5.892307(0)(4) \\
17 & -7.374398(1)(3) & -4.650356(1)(2) & -5.367820(0)(4) & -5.891210(0)(4) \\
18 & -7.128523(0)(6) & -4.632985(1)(4) & -5.341862(0)(4) & -5.857640(0)(2) \\
19 & -6.974804(0)(4) & -4.631229(0)(1) & -5.339562(0)(4) & -5.826138(0)(1) \\
20 & -6.910831(0)(4) & -4.627777(2)(2) & -5.338550(0)(2) & -5.819339(1)(4) \\
21 & -6.776769(2)(3) & -4.613176(1)(4) & -5.306673(1)(2) & -5.801340(0)(1) \\
22 & -6.759607(0)(3) & -4.606503(0)(4) & -5.301164(1)(4) & -5.783725(1)(4) \\
23 & -6.735370(1)(6) & -4.602550(2)(1) & -5.220173(1)(4) & -5.760944(1)(4) \\
24 & -6.671460(1)(6) & -4.594634(2)(4) & -5.219546(1)(4) & -5.674710(1)(2) \\
25 & -6.659101(1)(4) & -4.567206(1)(4) & -5.173984(1)(2) & -5.638308(0)(2) \\
26 & -6.632900(1)(1) & -4.504516(0)(2) & -5.168376(0)(1) & -5.617886(1)(2) \\
27 & -6.587695(1)(4) & -4.485153(1)(2) & -5.165203(1)(4) & -5.610840(0)(1) \\
28 & -6.584650(0)(8) & -4.466332(2)(1) & -5.109319(0)(1) & -5.607030(1)(4) \\
29 & -6.515256(0)(6) & -4.462375(1)(4) & -5.108918(0)(4) & -5.589100(1)(4) \\
30 & -6.495976(2)(6) & -4.454945(1)(4) & -5.106297(2)(1) & -5.523017(1)(4) \\
31 & -6.452904(2)(3) & -4.421657(1)(4) & -5.099047(2)(2) & -5.510477(0)(4) \\
32 & -6.447382(0)(4) & -4.373617(0)(1) & -5.078950(1)(4) & -5.499922(2)(1) \\
33 & -6.386656(1)(8) & -4.328330(1)(4) & -5.073168(0)(4) & -5.480655(0)(4) \\
34 & -6.371225(2)(6) & -4.307419(2)(4) & -5.065864(2)(4) & -5.473585(1)(1) \\
35 & -6.290827(1)(4) & -4.303792(0)(2) & -5.063913(1)(4) & -5.455337(2)(1) \\
36 & -6.271618(1)(2) & -4.301528(2)(4) & -5.058290(1)(4) & -5.446136(1)(4) \\
37 & -6.269614(1)(8) & -4.293005(1)(1) & -5.016248(0)(1) & -5.436872(0)(1) \\
38 & -6.263645(2)(8) & -4.286496(0)(4) & -5.001478(0)(4) & -5.436555(2)(2) \\
39 & -6.231479(2)(1) & -4.280238(0)(4) & -4.995843(0)(2) & -5.425527(1)(1) \\
40 & -6.229928(1)(8) & -4.273400(1)(1) & -4.994059(1)(1) & -5.405521(1)(4) \\
41 & -6.195617(2)(8) & -4.254363(2)(2) & -4.990596(1)(1) & -5.400260(1)(4) \\
42 & -6.176941(2)(8) & -4.252526(0)(4) & -4.981097(1)(2) & -5.394346(2)(4) \\
43 & -6.173178(1)(6) & -4.245270(2)(1) & -4.940270(0)(4) & -5.388913(1)(2) \\
44 & -6.148869(3)(4) & -4.244846(2)(1) & -4.930326(1)(4) & -5.352229(0)(2) \\
45 & -6.110109(1)(3) & -4.244222(2)(2) & -4.917262(1)(4) & -5.324424(0)(4) \\
\hline
\end{tabular}
\caption{Distinct energy levels of $H$ for $L=4$, listed in ascending order at four different coupling values. The notation follows that of \cref{tab:L2evals}.}
\label{tab:L4evals}
\end{table*}

\begin{table*}[h!]
\centering
\renewcommand{\arraystretch}{1.2}
\setlength{\tabcolsep}{6pt}
\begin{tabular}{|c|cc|cc|cc|cc|}
\hline
\multirow{2}{*}{Projection State} & \multicolumn{2}{c|}{(J=0, J'=1)} & \multicolumn{2}{c|}{(J=1, J'=0.5)} & \multicolumn{2}{c|}{(J=1, J'=0.66)} & \multicolumn{2}{c|}{(J=1, J'=0.75)} \\
\cline{2-9}
  & $(a)$ & $\langle \chi_a|\phi_1\rangle$ & $(a)$ & $\langle \chi_a|\phi_1\rangle$ & $(a)$ & $\langle \chi_a|\phi_1\rangle$ & $(a)$ & $\langle \chi_a|\phi_1\rangle$ \\
\hline
\multirow{30}{*}{$|\psi_5\rangle$} & 4 & -0.000523 & 2 & -2.103814 & 5 & -5.206269 & 5 & -0.674330 \\
  & 4 & -0.000417 & 2 & +3.151237 & 5 & -3.555676 & 5 & +0.113756 \\
  & 4 & +0.000049 & 2 & +8.303796 & 5 & -0.898546 & 5 & +0.382555 \\
  & 4 & -0.000396 & 2 & +1.254938 & 5 & +0.179952 & 5 & -0.342496 \\
  & 5 & +0.000208 & 3 & -1.583050 & 7 & -6.407277 & 6 & -0.281432 \\
  & 5 & -0.000080 & 3 & +6.913795 & 7 & +0.422894 & 6 & -1.030326 \\
  & 5 & -0.000112 & 3 & -2.804170 & 7 & +0.680648 & 6 & +1.642171 \\
  & 5 & -0.000163 & 3 & +5.180237 & 7 & +0.830089 & 6 & +0.604951 \\
  & 6 & +0.000176 & 11 & +0.030649 & 15 & +1.920424 & 15 & -3.035473 \\
  & 6 & -0.000298 & 11 & +0.084143 & 15 & +2.589075 & 15 & -1.346833 \\
  & 6 & +0.000825 & 11 & +0.028422 & 15 & -1.037317 & 15 & +2.433828 \\
  & 6 & +0.000158 & 11 & -0.009558 & 15 & -0.542344 & 15 & +0.066491 \\
  & 6 & +0.000392 & 18 & +1.201394 & 16 & -0.680921 & 20 & -0.224009 \\
  & 6 & +0.000159 & 18 & +1.229969 & 16 & -3.302649 & 20 & +2.456945 \\
  & 15 & -0.001336 & 18 & -3.925980 & 16 & +1.427879 & 20 & +1.640985 \\
  & 15 & -0.002694 & 18 & -2.040563 & 16 & +1.654346 & 20 & +1.696154 \\
  & 15 & -0.001904 & 21 & +3.598485 & 22 & -0.012883 & 22 & +0.110409 \\
  & 15 & -0.001241 & 21 & +1.199243 & 22 & -0.748783 & 22 & +2.470518 \\
  & 15 & +0.000172 & 21 & -2.574054 & 22 & -0.412948 & 22 & -1.564518 \\
  & 15 & -0.001100 & 21 & +1.990788 & 22 & +0.876721 & 22 & +2.555769 \\
  & 16 & +0.001002 & 29 & -1.803682 & 23 & +0.035347 & 23 & +1.701200 \\
  & 16 & -0.001189 & 29 & -0.794814 & 23 & -0.180211 & 23 & +2.251446 \\
  & 16 & +0.000731 & 29 & -0.045448 & 23 & +2.228450 & 23 & -0.354720 \\
  & 16 & +0.000611 & 29 & -0.553695 & 23 & +0.831433 & 23 & -2.055897 \\
  & 16 & -0.001217 & 31 & -0.083550 & 24 & +0.232138 & 28 & -0.487518 \\
  & 16 & -0.001145 & 31 & +1.434149 & 24 & -0.385640 & 28 & -0.109098 \\
  & 16 & +0.000989 & 31 & +0.537135 & 24 & -0.088382 & 28 & +0.421335 \\
  & 16 & +0.000466 & 31 & -1.359334 & 24 & -0.515504 & 28 & +0.208351 \\
  & 23 & -0.001943 & 33 & +0.952714 & 32 & -1.168784 & 30 & -0.371020 \\
  & 23 & +0.000012 & 33 & -0.973504 & 32 & +3.279464 & 30 & -1.407949 \\
  & 23 & +0.002799 & 33 & +0.208153 & 32 & +0.171437 & 30 & -1.945576 \\
  & 23 & -0.000229 & 33 & +0.721085 & 32 & -0.066812 & 30 & -0.642030 \\
\hline
\end{tabular}
\caption{Overlap of the projection state $|\phi_1\rangle$ in the $Q=-1$ projection subspace with the lowest distinct energy eigenstates $|\chi_a\rangle$ (up to some high energy cutoff) at the different couplings.}\label{tab:L4ov-phi}
\end{table*}

\begin{table*}[h!]
\centering
\renewcommand{\arraystretch}{1.2}
\setlength{\tabcolsep}{6pt}
\smaller
\begin{tabular}{|c|cc|cc|cc|cc|}
\hline
\multirow{2}{*}{Reference State} & \multicolumn{2}{c|}{(J=0, J'=1)} & \multicolumn{2}{c|}{(J=1, J'=0.5)} & \multicolumn{2}{c|}{(J=1, J'=0.66)} & \multicolumn{2}{c|}{(J=1, J'=0.75)} \\
\cline{2-9}
  & $(a)$ & $\langle\chi_a|\psi_i\rangle$ & $(a)$ & $\langle\chi_a|\psi_i\rangle$  & $(a)$ & $\langle\chi_a|\psi_i\rangle$  & $(a)$ & $\langle\chi_a|\psi_i\rangle$  \\
\hline
\multirow{16}{*}{$|\psi_{1}\rangle$} & 1 & -73.572245 & 6 & +1.029357 & 2 & +51.797366 & 1 & -58.563070 \\
  & 2 & -81.545673 & 32 & +28.703482 & 9 & +17.469325 & 3 & -70.038105 \\
  & 2 & -81.545673 & 44 & -0.270035 & 14 & -67.036584 & 3 & +70.038105 \\
  & 3 & +60.170614 & 44 & -0.270035 & 14 & +67.036584 & 8 & +17.009229 \\
  & 3 & -73.693652 & 44 & +0.220483 & 37 & -38.647933 & 27 & +35.546385 \\
  & 3 & -73.693652 &  &  &  &  & 32 & -62.463940 \\
  & 9 & -57.258901 &  &  &  &  & 32 & +51.001593 \\
  & 9 & -57.258901 &  &  &  &  & 32 & +62.463940 \\
  & 9 & -44.352554 &  &  &  &  &  &  \\
  & 9 & +44.352554 &  &  &  &  &  &  \\
  & 12 & +32.818994 &  &  &  &  &  &  \\
  & 26 & -40.137590 &  &  &  &  &  &  \\
  & 26 & +40.137590 &  &  &  &  &  &  \\
  & 39 & +11.323668 &  &  &  &  &  &  \\
  & 39 & +11.323668 &  &  &  &  &  &  \\
  & 39 & +9.245737 &  &  &  &  &  &  \\
\hline
\multirow{7}{*}{$|\psi_{2}\rangle$} & 1 & -0.887215 & 6 & +0.007772 & 2 & +0.558333 & 1 & -0.647766 \\
  & 3 & +1.062861 & 32 & +0.277715 & 9 & +0.205151 & 8 & +0.232299 \\
  & 12 & +0.831569 & 44 & +0.004085 & 37 & -0.533194 & 27 & +0.533529 \\
  & 31 & -0.029389 &  &  &  &  & 32 & +0.848457 \\
  & 31 & +1.575769 &  &  &  &  &  &  \\
  & 31 & -1.290705 &  &  &  &  &  &  \\
  & 39 & +0.771978 &  &  &  &  &  &  \\
\hline
\multirow{12}{*}{$|\psi_{3}\rangle$} & 1 & -4.771500 & 1 & -0.250000 & 1 & -0.250000 & 1 & -6.014836 \\
  & 7 & -6.843183 & 4 & +0.000515 & 2 & +6.113502 & 2 & +0.776693 \\
  & 7 & -3.955590 & 4 & +0.173052 & 3 & +0.113003 & 2 & -4.397768 \\
  & 7 & -2.793187 & 6 & -0.681085 & 3 & -2.694568 & 7 & +0.092138 \\
  & 12 & -4.663593 & 7 & +1.146214 & 6 & -0.995001 & 8 & +0.289489 \\
  & 22 & -2.263543 & 26 & +4.154496 & 9 & +2.016729 & 10 & +0.250000 \\
  & 22 & -2.751587 & 26 & -0.661060 & 20 & -6.759096 & 18 & +2.428977 \\
  & 22 & -0.465120 & 32 & +7.003587 & 20 & +0.652893 & 18 & -5.466307 \\
  &  &  & 35 & -1.599671 & 26 & -3.417394 & 21 & -3.906933 \\
  &  &  & 35 & -3.784518 & 37 & +3.184226 & 27 & -3.945731 \\
  &  &  &  &  &  &  & 44 & -0.909664 \\
  &  &  &  &  &  &  & 44 & -0.085537 \\
\hline
\multirow{15}{*}{$|\psi_{4}\rangle$} & 1 & -0.253664 & 1 & -0.250000 & 1 & -0.250000 & 1 & -0.381669 \\
  & 3 & -0.448050 & 4 & +0.000920 & 2 & +0.404230 & 2 & -0.015514 \\
  & 7 & +0.574956 & 4 & +0.309110 & 3 & +0.003757 & 2 & +0.087845 \\
  & 7 & +0.332344 & 6 & -0.332405 & 3 & -0.089582 & 7 & -0.224484 \\
  & 7 & +0.234680 & 7 & +0.003566 & 6 & -0.085798 & 8 & +0.095751 \\
  & 12 & -0.411226 & 20 & -0.011440 & 9 & +0.320786 & 10 & +0.250000 \\
  & 21 & +0.426513 & 20 & +0.241323 & 20 & +0.660340 & 18 & -0.283768 \\
  & 21 & +0.046423 & 26 & -0.453246 & 20 & -0.063785 & 18 & +0.638608 \\
  & 21 & +0.970522 & 26 & +0.072120 & 26 & +0.438696 & 21 & +0.386064 \\
  & 22 & -0.206410 & 32 & +0.788283 & 31 & +0.081155 & 27 & -0.739562 \\
  & 22 & -0.250914 & 35 & +0.086646 & 31 & -0.276906 & 32 & -0.337693 \\
  & 22 & -0.042414 & 35 & +0.204989 & 37 & +0.702255 & 38 & +0.008725 \\
  & 39 & +0.052647 & 43 & +0.186901 &  &  & 38 & -0.461112 \\
  &  &  & 44 & -0.343425 &  &  & 44 & -0.418365 \\
  &  &  &  &  &  &  & 44 & -0.039339 \\
\hline
\end{tabular}
\caption{Overlaps of the four projection states that span the $Q=+1$ projection subspace with the lowest distinct eigenstates $|\chi_a\rangle$ (up to some high energy cutoff) at the different couplings.}\label{tab:L4ov-psi}
\end{table*}

\clearpage

\bibliographystyle{apsrev4-2}
\showtitleinbib
\bibliography{apsfix,ref,refE8}

\end{document}